\documentclass{siamart1116}
\usepackage{amsfonts}

\renewcommand\Re{\operatorname{Re}}

\title{Localized Patterns in Periodically Forced Systems: \\
  II. Patterns with Non-Zero Wavenumber}

\author{A. S. Alnahdi%
  \thanks{Department of Mathematics \& Statistics, College of
    Science, Al Imam Mohammad Ibn Saud Islamic University, PO Box
    240455, Riyadh 11322, KSA (\email{asalnahdi@imamu.edu.sa}).}
  \and
  J. Niesen%
  \thanks{School of Mathematics, University of Leeds, Leeds, LS2 9JT,
    UK (\email{j.niesen@leeds.ac.uk, a.m.rucklidge@leeds.ac.uk}).}
  \and
  A. M. Rucklidge\footnotemark[2]}

\begin{document}
\bibliographystyle{siam}
\maketitle

\begin{abstract}
  In pattern-forming systems, localized patterns are readily found
  when stable patterns exist at the same parameter values as the
  stable unpatterned state.  Oscillons are spatially localized,
  time-periodic structures, which have been found experimentally in
  systems that are driven by a time-periodic force, for example, in
  the Faraday wave experiment. This paper examines the existence of
  oscillatory localized states in a PDE model with single frequency
  time dependent forcing, introduced in~\cite{RS} as a
  phenomenological model of the Faraday wave experiment.  We choose
  parameters so that patterns set in with non-zero wavenumber (in
  contrast to~\cite{ANRW}). In the limit of weak damping, weak
  detuning, weak forcing, small group velocity, and small amplitude,
  we reduce the model PDE to the coupled forced complex
  Ginzburg--Landau equations.  We find localized solutions and snaking
  behaviour in the coupled forced complex Ginzburg--Landau equations
  and relate these to oscillons that we find in the model PDE. Close
  to onset, the agreement is excellent. The periodic forcing for the
  PDE and the explicit derivation of the amplitude equations make our
  work relevant to the experimentally observed oscillons.
\end{abstract}


\pagestyle{myheadings}
\thispagestyle{plain}
\headers{A S Alnahdi,  J Niesen, and A M Rucklidge}%
{Localized Patterns in Periodically Forced Systems II}

\noindent\makebox[\linewidth]{\rule{\textwidth}{1pt}}

\section{Introduction}

\begin{figure}
  \centering
  \includegraphics[height=60mm]{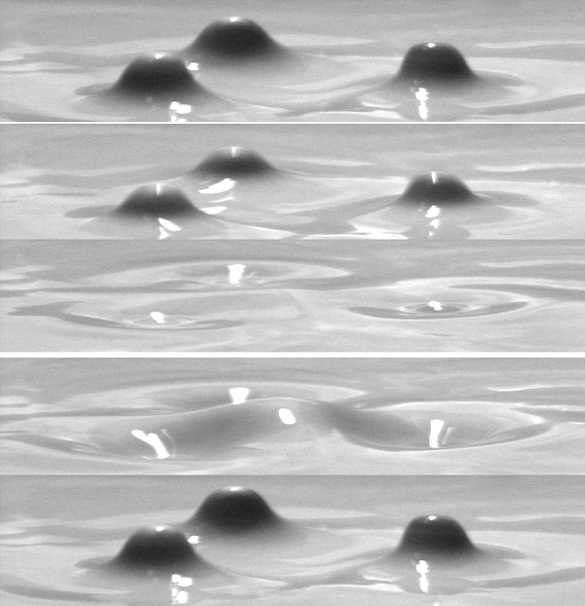}\hspace{10mm}
  \includegraphics[height=60mm]{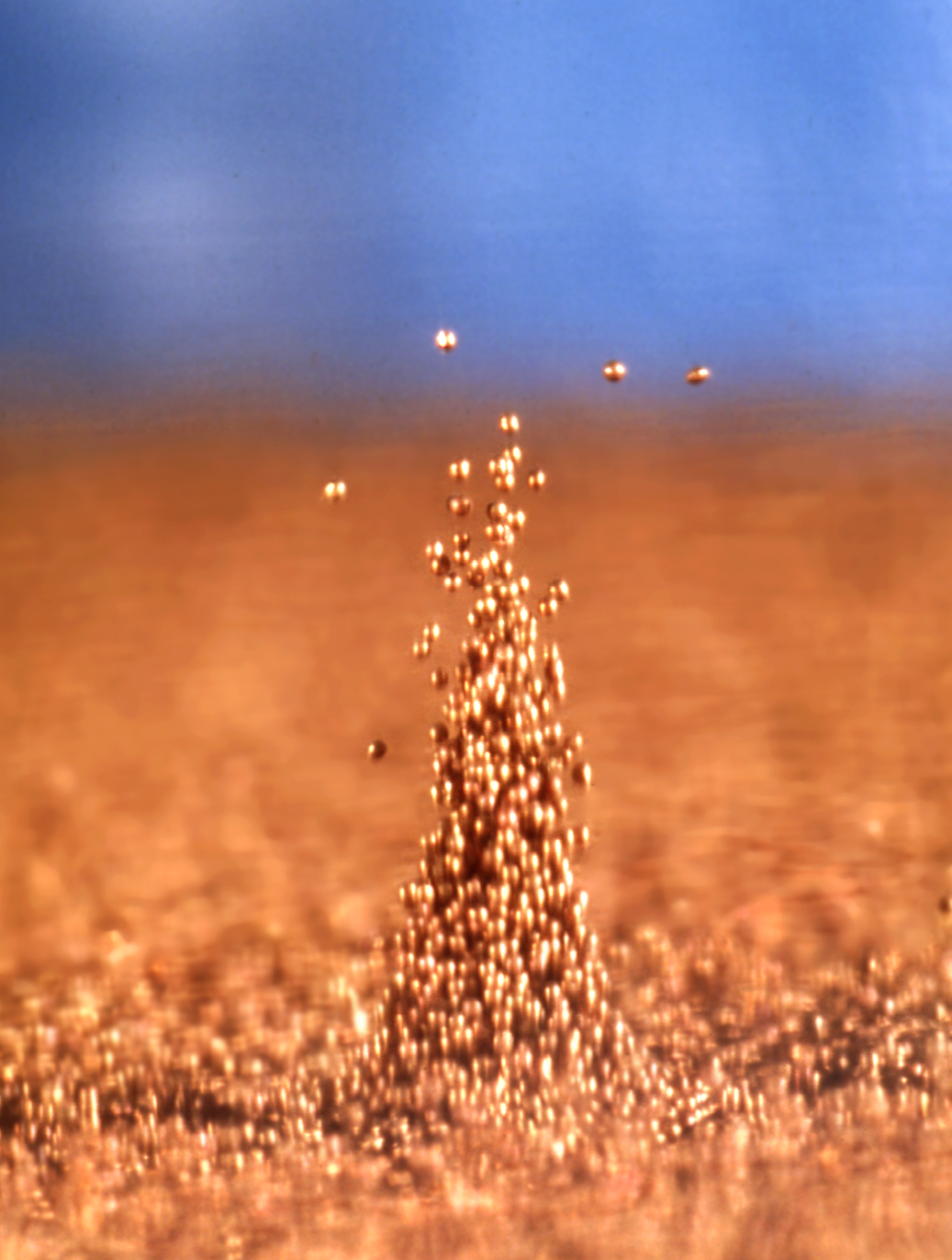}
  \caption{Left: A triad of oscillons in a vertically vibrated
    colloidal suspension, with time running from top to bottom (taken
    from~\cite{LHARF}). Right: An oscillon in a vertically vibrated
    layer of bronze beads (courtesy of Paul Umbanhowar, Northwestern
    University).}
  \label{fig:experiment}
\end{figure}

Spatially localized structures are common in pattern forming systems,
appearing in fluid mechanics, chemical reactions, optics and granular
media~\cite{D2, K}.  Much progress has been made on the analysis of
steady problems, where bistability between a steady pattern and the
zero state leads to steady localized patterns bounded by stationary
fronts between these two states \cite{BK2, D1}. In contrast,
oscillons, which are oscillating localized structures in a stationary
background in periodically forced dissipative systems, are relatively
less well understood. Oscillons have been found experimentally in
fluid surface wave experiments \cite{AF, EF, LAF, LHARF, SXP, WKR},
chemical reactions \cite{PQS}, optical systems~\cite{L}, and vibrated
granular media problems~\cite{BSSMS, TA, UMS}. In the surface wave
experiments (see the left panel of \cref{fig:experiment}), the fluid
container is driven by vertical vibrations. When these are strong
enough, the surface of the system becomes unstable (the Faraday
instability) \cite{F}, and standing waves are found on the surface of
the fluid. Oscillons have been found when this primary bifurcation is
subcritical \cite{CR}, and these take the form of alternating conical
peaks and craters against a stationary background.  A second striking
example of oscillons was found in a vertically vibrated thin layer of
granular particles~\cite{UMS}, as depicted in the right panel of
\cref{fig:experiment}.  As with the surface wave experiments,
oscillons take the shape of alternating peaks and craters.  The
observation of oscillons in these experiments has motivated our
theoretical investigation into the existence of these states and their
stability in a model PDE with explicit time-dependent forcing. In both
of these experiments, the forcing (vertical vibration) is
time-periodic with frequency~$2\Omega$, and the oscillons themselves
vibrate with either the same frequency~($2\Omega$) as the forcing
(harmonic) or with half the frequency~($\Omega$) of the forcing
(subharmonic). We focus on the subharmonic case, because this is the
most relevant for single-frequency forcing as considered here; in
contrast, harmonic oscillations play an important role in the presence
of multi-frequency forcing~\cite{TS}.

A subharmonic standing wave modulated slowly in time is described by
an ansatz of the form
\begin{equation}
\label{eq:sw-ansatz}
U(t,x) = A(T) e^{i \Omega t} \cos(kx) + \text{c.c.}, 
\end{equation}
for a real scalar variable $U$ depending on a (fast) time variable~$t$
and a spatial variable~$x$. Here, $A$~is a complex amplitude depending
on a slow time scale~$T$; also, $k$~is the wavenumber and c.c.~stands
for complex conjugate. Phase shifts in~$A$ correspond to translations
in time. Symmetry considerations then lead to an amplitude equation of
the form
\begin{equation}
\label{eq:gl-no-diffusion}
A_T = (\rho + i\nu) A + C |A|^2 A + i \Gamma \bar{A},
\end{equation}
where the real parameter $\Gamma$ describes the strength of the
forcing. The parameters $\rho$ and~$\nu$ are real but $C$ is
complex. The last term (with~$\bar{A}$) breaks the phase symmetry
of~$A$ and thus the corresponding time-translation symmetry: the phase
of~$A$ is not arbitrary becuase the forcing in the original system is
time dependent. The factor~$i$ in the last term can be removed by
applying a phase shift.  See~\cite{DC} for a discussion of this (and
related) amplitude equations.

In the case of spatially localized oscillons, we also have to include
spatial modulations, so that the amplitude~$A$ in~\cref{eq:sw-ansatz}
depends not only on~$T$ but also on a slow spatial variable~$X$. It
would seem logical that this ansatz would lead to a diffusion term
to~\cref{eq:gl-no-diffusion}, yielding a forced complex
Ginzburg--Landau (FCGL) equation which is typically written down
without derivation~\cite{DL, MBK, MS, BYK1}:
\begin{equation}
  \label{eq:fcgl}
  A_T = (\rho + i \nu) A - 2 (\alpha + i \beta) A_{XX} + C |A|^2 A 
  + i \Gamma \bar{A}.
\end{equation}
Here, $\alpha$ and~$\beta$ are real parameters; the factor~$-2$ is
included for comparison with the results that we will derive in this
paper.  Burke, Yochelis and Knobloch~\cite{BYK} showed that this
equation admits localized solutions. In~\cite{ANRW}, the FCGL equation
was derived from a model PDE in which patterns are formed with zero
wavenumber at onset; the agreement between the localized solutions in
the model PDE and those in~\cref{eq:fcgl} was excellent.

However, in the Faraday wave experiment, the preferred wavenumber is
non-zero at onset~\cite{BU}.  Nevertheless, the FCGL equation has
sometimes been used as an amplitude equation for Faraday wave and
granular oscillons \cite{AT, DL, TA, ZV2}. In this paper, we argue
that this is not appropriate; instead, a system of two coupled forced
complex Ginzburg--Landau equations should be used, as was done
in~\cite{MKV, RI}.

In order to demonstrate explicitly the origin and correctness of the
coupled FCGL equations as amplitude equations for oscillons, we use a
PDE model with single-frequency time-dependent forcing, introduced in
\cite{RS} as phenomenological model of the Faraday wave experiment.
We simplify the PDE by removing quadratic terms, and by taking the
parametric forcing to be $\cos(2 t)$, where $t$ is the fast time
scale. The resulting model PDE is then
\begin{equation}\label{eq:rs2}
  U_{t} = (\mu + i\omega) U + ( \alpha + i \beta) U_{xx}
  + (\gamma + i \delta) U_{xxxx} + C |U|^2 U + i \Re{(U)} F \cos(2t), 
\end{equation}
where $U(x,t)$ is a complex function, $\mu<0$ is the distance from
onset of the oscillatory instability, $\omega$, $\alpha$, $\beta$,
$\gamma$, $\delta$ and $F$ are real parameters, and $C$ is a complex
parameter. The $\cos(2t)$ term makes this PDE non-autonomous.  In this
model, the dispersion relation can be readily controlled so the
wavenumber at onset can be chosen be zero or non-zero, and the
nonlinear terms are chosen to be simple in order that the weakly
nonlinear theory and numerical solutions can be computed easily.
In~\cite{ANRW}, the wavenumber at the onset of pattern formation was
zero, and the FCGL equation was derived as a description of the
localized solution. There, we did not require the fourth-order
derivatives in~\cref{eq:rs2}. In contrast, in the current study we use
the dispersion relation to set the wavenumber to be~1 at onset, and
therefore we need to retain the term $(\gamma+i\delta)U_{xxxx}$ with
the fourth-order spatial derivatives.

Our aim is to find and analyze spatially localized oscillons with
non-zero wavenumber in the PDE model \cref{eq:rs2} theoretically and
numerically in 1D and numerically in~2D. The approach will be similar
to that in \cite{ANRW}, though conceptionally more complicated since
we have to consider the interaction between left- and right-travelling
waves and the effect of a non-zero group velocity, leading to coupled
amplitude equations.  Although we will work with a model PDE, our
approach will show how localized solutions might be studied in PDEs
more directly connected to the Faraday wave experiment, such as the
Zhang--Vi\~nals model \cite{ZV}, and how weakly nonlinear calculations
from the Navier--Stokes equations~\cite{SG} might be extended to the
oscillons observed in the Faraday wave experiment.

In this case we can model waves with a slowly varying envelope in one
spatial dimension by looking at solutions of the form
\begin{equation}\label{eq:sol1}
  U(x,t) = A(X,T) e^{i(t+x)} + B(X,T) e^{i(t-x)},
\end{equation}
where $X$ and $T$ are slow scales, and $x$ and $t$ are scaled so that
the wave has critical wavenumber $k_c=1$ and critical frequency
$\Omega_c=1$. Commonly the complex conjugate is added to an ansatz of
the form~\cref{eq:sol1} in order to make $U$ real, but our
PDE~\cref{eq:rs2} admits complex solutions (we argue in the conclusion
that this does not make a material difference).  In order to cover the
symmetries of the PDE model, we include both the left- and
right-travelling waves (with amplitudes~$A$ and $B$, respectively) but
the time dependence will be $e^{it}$ only, without $e^{-it}$. In
\cref{lineartheory_chap3}, we explain in detail how the solution of
the linear operator, which we will define later, involves $e^{it}$
only. The $+1$ frequency dominates at leading order because of our
choice of dispersion relation. Here, we will focus primarily on the
one-dimensional case. Two-dimensional localized oscillons are
discussed briefly at the end and studied numerically in more detail
in~\cite{A}.

We start by showing some numerical examples of oscillons in the model
PDE~\cref{eq:rs2} and bifurcation diagrams exhibiting snaking, where
branches of solutions go back and forth as parameters are varied and
the width of the localized pattern increases. We will
do an asymptotic reduction of the model PDE to the coupled FCGL
equations in the limit of weak damping, weak detuning, weak forcing,
small group velocity, and small amplitude, and we will study the
properties of the coupled FCGL equations. Some numerical examples of
spatially localized oscillons in the coupled FCGL equations will be
given. We will also investigate the effect of changing the group
velocity. Furthermore, we will reduce the coupled FCGL equations to
the real Ginzburg--Landau equation in a further limit of weak forcing
and small amplitude close to onset. The real Ginzburg--Landau equation
has exact localized sech solutions. Throughout, we will use weakly
nonlinear theory by introducing a multiple scale expansion to do the
reduction to the amplitude equations. We conclude with numerical
examples of strongly localized oscillons in 1D and~2D.

\section{Numerical results for the model PDE}
\label{sec:numerics}

\begin{figure}
  \centering
  \includegraphics[width=\textwidth]{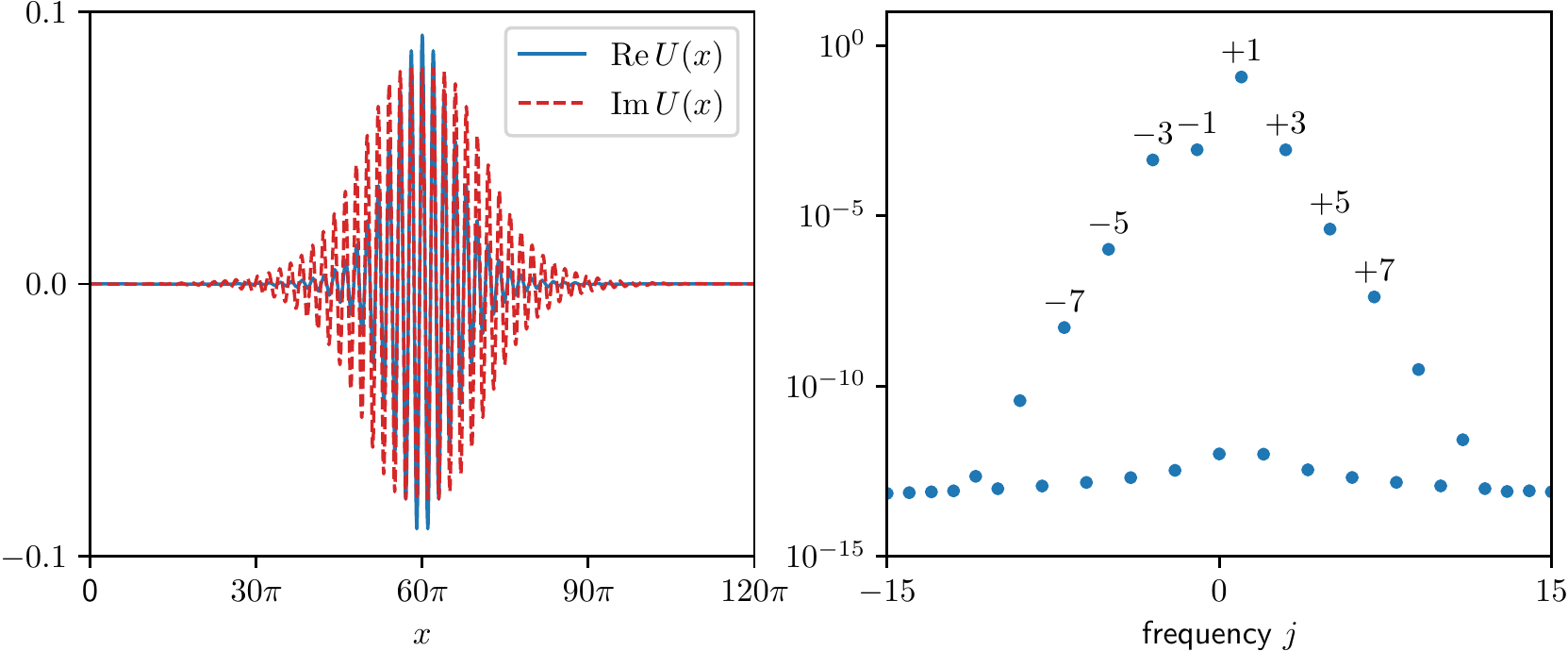}
  \caption{Left: Stable oscillon solution of \cref{eq:rs2} found by
    time-stepping, with $\mu = -0.255$, $\omega = 1.5325$,
    $\alpha = -0.5$, $\beta = 1$, $\gamma = -0.25$, $\delta = 0.4875$,
    $C = -1-2.5i$, and $F = 0.0585$. The solution is plotted at
    $t=0$. Right: Amplitude of the $e^{ijt}$ mode with frequency~$j$
    when expanding the solution in the left panel at $x = 60\pi$ as a
    Fourier series in time: the frequency $+1$ component is the
    strongest, followed by frequencies $-3$, $-1$ and $+3$, as
    expected, with the other frequencies at least two orders of
    magnitude weaker.}
  \label{fig:ex_pde_chap3_timestepping}
\end{figure}

Similar to the methodology that was used in \cite{ANRW}, we present
numerical simulations of the PDE model \cref{eq:rs2} by time-stepping
and continuation. The choice of parameters is guided by the asymptotic
analysis in the remainder of the paper: all modes are damped in the
absence of forcing, but the modes with wavenumber $k \simeq \pm 1$
are only weakly damped, the forcing is also weak, and the group
velocity is small.  We discretize the PDE using a Fourier
pseudospectral method and the resulting system of ODEs is solved with
a fourth-order exponential time differencing (ETD) method~\cite{CM}.
Most experiments are done on a domain of size $L = 120\pi$
(60~wavelengths), in which case we use 2048~grid points. Solving the
PDE from an appropriate initial condition, we find the localized
solution plotted in the left panel of
\cref{fig:ex_pde_chap3_timestepping}.

To do continuation from this localized solution, we represent
solutions by a truncated Fourier series in time with frequencies $-3$,
$-1$, $1$ and~$3$.  The choice of these frequencies comes from the
choice of parameters: the linearized PDE at wavenumber~$\pm1$ looks
like $\frac{\partial u}{\partial t}=iu$ (writing $U = u(t) e^{ix}$),
so the strongest Fourier component of~$u$ looks like $e^{it}$; then
putting $u=e^{it}$ into the forcing $\Re(e^{it})\cos(2t)$ generates
the frequencies $-3$, $-1$, $1$ and~$3$, as described
in~\cite{ANRW}. We also checked numerically that the frequencies
$\pm1$ and~$\pm3$ dominate (see the right panel of
\cref{fig:ex_pde_chap3_timestepping}).

\begin{figure}
  \centering
  \includegraphics[width=\textwidth]{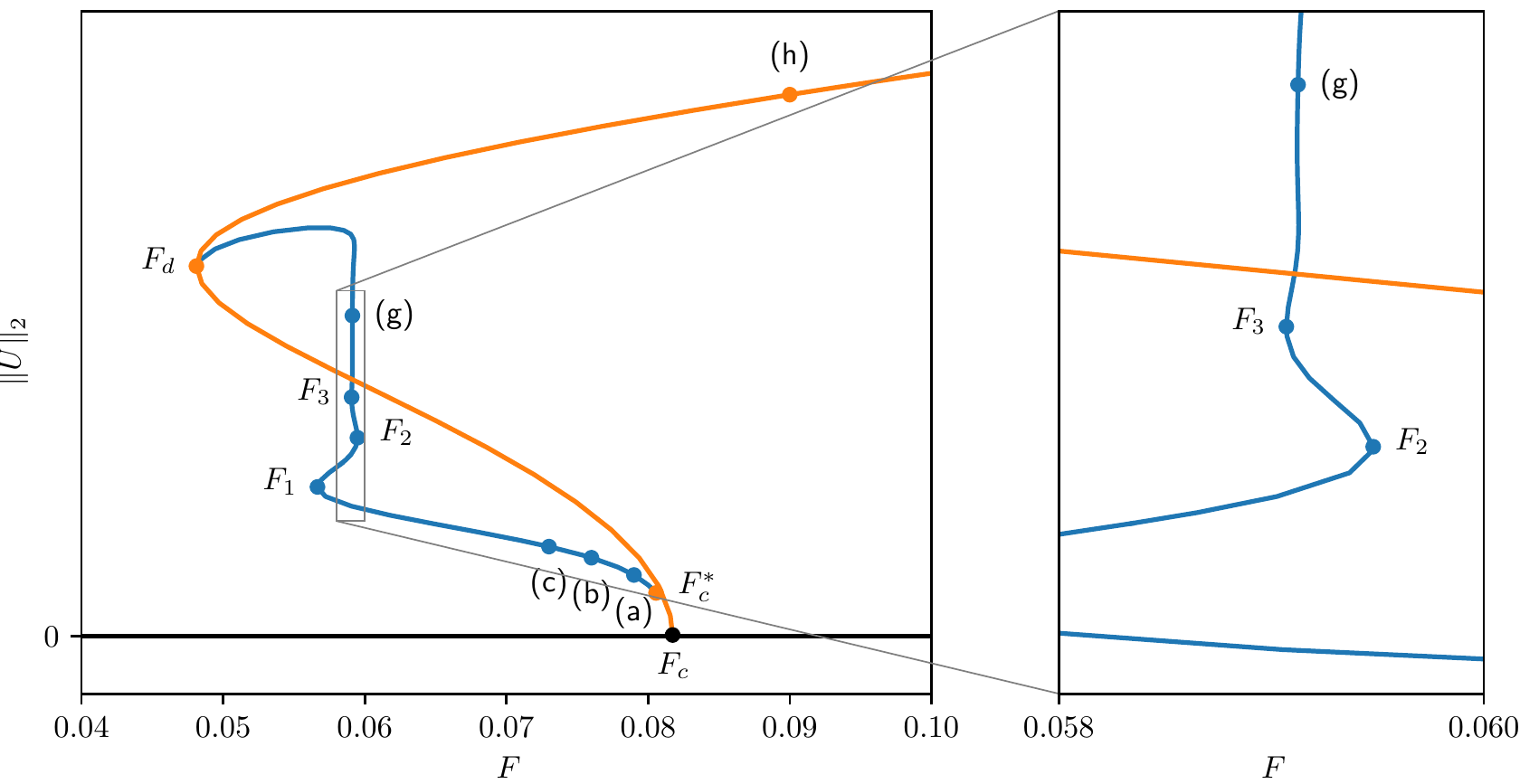}
  \caption{Bifurcation diagram of \cref{eq:rs2} in the weak damping
    limit in a domain of size $L_x=120\pi$ with parameters as in
    \cref{fig:ex_pde_chap3_timestepping}. The branch with periodic
    solutions is plotted in red. The bistability region is between
    $F_d=0.04811$ and $F_c=0.08173$. The branch with localized
    solutions (blue) starts at $F^* _c = 0.08056$ and has folds at
    $F_1 = 0.05666$, $F_2 = 0.05948$ and $F_3 = 0.05912$. Solutions at
    (a), (b), (c), $F_1$, $F_2$, $F_3$, (g) and~(h) are shown in
    \cref{fig:five_oscillons}.}
  \label{fig:bif_pde}
\end{figure}

The bifurcation diagram of \cref{eq:rs2} as computed by
AUTO~\cite{Doe} is given in \cref{fig:bif_pde}. The subcritical
transition from the zero state to the pattern occurs at the
bifurcation point $F_c = 0.08173$.  The saddle-node point where the
unstable periodic pattern becomes stable is at $F_d = 0.04811$. The
bistability region where we look for the branch of localized states is
between $F_c$ and $F_d$. The branch of localized solutions bifurcates
from the branch of periodic patterns at $F_c^* = 0.08056$, which is
away from $F_c$ because of the finite domain. Stable localized
solutions are located between $F_1 = 0.05666$ and $F_2 = 0.05948$.

\begin{figure}
  \centering
  \includegraphics[width=\textwidth]{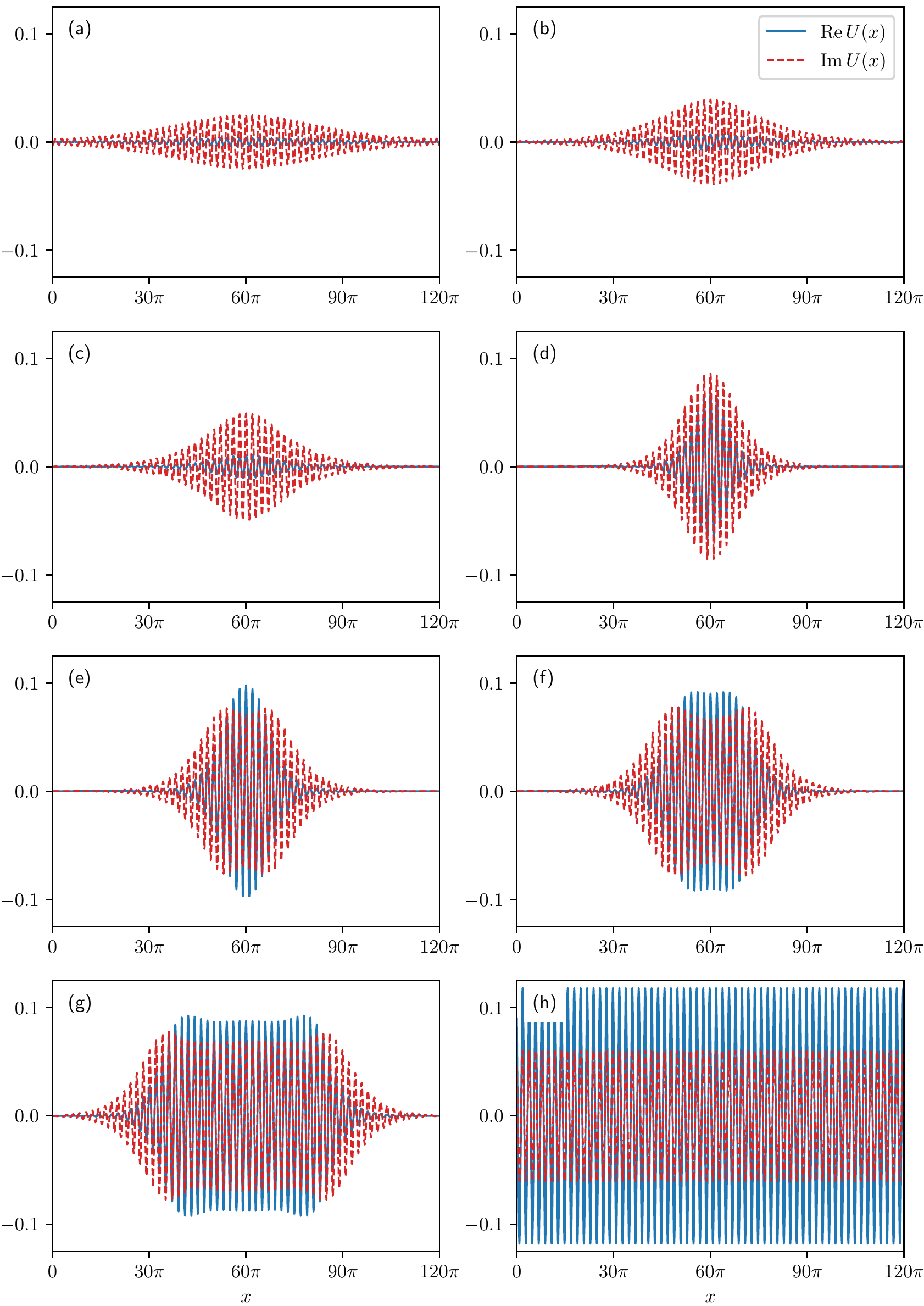}
  \caption{Solutions along the branch of localized solutions in the
    bifurcation diagram in \cref{fig:bif_pde}, at (a)~$F = 0.079$,
    (b)~$F = 0.076$, (c)~$F = 0.073$, (d)~the fold at~$F_1 = 0.05666$,
    (e)~the fold at~$F_2 = 0.05948$, (f)~the fold at~$F_3 = 0.05912$,
    and the point~(g). Solution~(h) is on the periodic branch at $F =
    0.09$.}
  \label{fig:five_oscillons}
\end{figure}

Examples of solutions along the branch of localized solutions in
\cref{fig:bif_pde} are given in \cref{fig:five_oscillons}.  Near the
point $F_c ^*$ where the branch of localized solutions bifurcates, the
localized solutions look like the periodic patterns: small amplitude
oscillations which are not very localized (see
\cref{fig:five_oscillons}(a)). As we go along the branch of localized
solutions, the amplitude increases and the unstable oscillons become
more localized (\cref{fig:five_oscillons}(b)--(c)).  At $F_1 =
0.05695$, the localized oscillons stabilize
(\cref{fig:five_oscillons}(d)) and then they lose stability again at
$F_2 = 0.05987$ (\cref{fig:five_oscillons}(e)) as the branch of
solutions snakes back and forth.  The next saddle-node point is at
$F_3 = 0.05912$ (\cref{fig:five_oscillons}(f)). It appears from the
numerical results that the parameter intervals between successive
saddle-node points shrinks to zero as we continue on the branch with
localized solutions; this is called \emph{collapsed snaking}
in~\cite{MBK}. However, we suspect that our numerics are misleading,
partially because the domain size is too small, and that in fact, the
odd and even saddle-node points asymptote to parameter values which
are close to each other but not equal.  The branch of localized
solution connects to the pattern branch close to the saddle-node point
$F_d$.  \Cref{fig:five_oscillons}(h) shows a typical periodic pattern.
All solutions in \cref{fig:bif_pde,fig:five_oscillons} satisfy $U(x,t)
= U(-x,t)$ for a suitably chosen origin. We have not found solutions
with any other symmetry.

In the remainder of the paper, we will analyze these oscillons and
derive an asymptotic expression for their amplitude, which will be
compared to the numerical solutions in
\cref{fig:compare_with_as}.

\section{Derivation of the coupled forced complex 
  Ginzburg--Landau (FCGL) equation}
\label{deriveFCCGL}

In this section we will study the PDE model \cref{eq:rs2} in the
limit of weak damping, weak detuning, weak forcing and small amplitude
in order to derive its amplitude equation. In addition, we will need
to assume that the group velocity is small. We start with linearizing
\cref{eq:rs2} about zero, and  we consider solutions of the
form $U(x,t) = e^{\sigma t + ikx}$, where $\sigma$ is the complex growth
rate of a mode with wavenumber~$k$. Without taking any limits and
without considering the forcing, the growth rate is given by
\begin{equation}
  \label{eq:growth3}
  \sigma = \mu - \alpha k^2 + \gamma k^4 + i (\omega - \beta k^2 + \delta k^4),
\end{equation} 
so $\sigma_r = \mu - \alpha k^2 + \gamma k^4$ gives the damping rate
of modes with wavenumber~$k$, and
$\sigma_i = \omega - \beta k^2 + \delta k^4$ gives the frequency of
oscillation.  We will also need the group velocity of the waves, which
is $d\sigma_i(k) / dk = -2 \beta k + 4 \delta k^3$.

We will choose parameters so that we are in a weak damping, weak
detuning, and small group velocity limit for modes with wavenumber
$k=1$. Specifically, in order to find spatially localized oscillons
and to do the reduction to the amplitude equation, we will impose the
following.

\paragraph{Stability in the absence of forcing}
To have waves with all wavenumbers linearly damped, we require that
$\sigma_r(k) < 0$, for all $k$. It follows that $\mu < 0$,
$\alpha > -2\sqrt{\mu\gamma}$ and $\gamma < 0$. With $\alpha < 0$ we
have a non-monotonic growth rate.

\paragraph{Preferred wavenumber}
We want the damping to be weakest for $k = \pm1$. Thus, we require
that the growth rate~$\sigma_r$ achieves a maximum when the wavenumber
$k$ is~1, so 
$\frac{d}{dk} \sigma_r(k=1) = -2 \alpha + 4 \gamma = 0$.  This gives
the condition $\alpha = 2\gamma$.

\begin{figure}
  \centering
  \includegraphics[width=\textwidth]{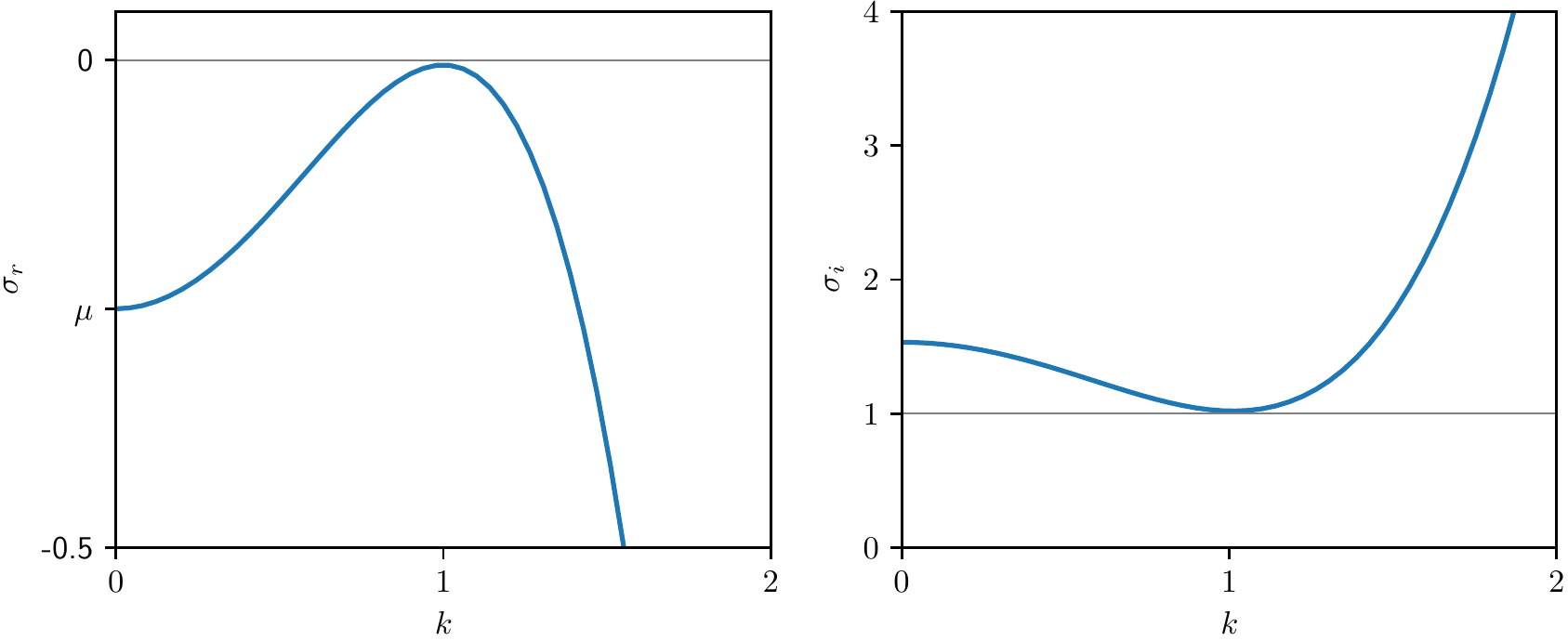}
  \caption{The growth rate (left panel) and dispersion relation (right
    panel) of equation \cref{eq:rs2} with $\mu = -0.255$,
    $\omega = 1.5325$, $\alpha = -0.5$, $\beta = 1$, $\gamma = -0.25$
    and $\delta = 0.4875$. In this case, the group velocity is small
    at $k=1$ because this is close to the minimum of the dispersion
    relation.}
  \label{fig:growth2}
\end{figure}

\paragraph{Weak damping}
We also need to make the growth rate $\sigma_r$ to be close to zero
when $k = \pm1$. Therefore, we introduce a small parameter
$\epsilon \ll 1$ and a new parameter $\rho$, so that we have
$\sigma_r(k=1) = \mu - \alpha + \gamma = \epsilon^2 \rho$, where
$\rho<0$. Thus, $\mu = \frac12 \alpha + \epsilon^2\rho$.
\Cref{fig:growth2}(a) shows an example of the real part of the growth
rate.

\paragraph{Weak detuning}
We want waves with $k \simeq \pm1$ to be subharmonically driven by
$\cos(2t)$, so the frequency of the oscillation~$\sigma_i$ should be
close to~$1$ at $k=1$. Therefore, we write $\sigma_i(k=1) = \omega -
\beta + \delta = 1 + \epsilon^2 \nu$, where $\nu$ is the detuning.

\paragraph{Small group velocity}
We require the group velocity $\frac{d\sigma_i}{dk} = -2 k \beta +
4\delta k^3$ to be~$O(\epsilon)$ at $k = \pm1$, so we have $-2 \beta +
4 \delta = \epsilon v_g$. This is needed to allow the group velocity
in the subsequent amplitude equations to appear at the same order as
all the other terms. We discuss the consequences of choosing a small
group velocity in \cref{sec:discussion}.  \Cref{fig:growth2}(b) shows
an example of the dispersion relation~$\sigma_i(k)$.

\paragraph{Weak forcing}
To perform the weakly nonlinear theory, we assume that the forcing is
weak, and so we scale the forcing amplitude to be $O(\epsilon^2)$,
writing $F = 4 \epsilon^2 \Gamma$.

\begin{table}
  \caption{Relationships between parameters 
    $(\mu, \omega, \alpha, \beta, \gamma, \delta, F)$ of the PDE model 
    and the parameters $(\rho, \nu, \alpha, \beta, v_g, \Gamma)$ of
    the coupled FCGL equations. Note that these relationships depend on the 
    choice of~$\epsilon$. The parameters $\alpha$ and $\beta$ are the 
    same in both models.}
  \label{tb:table1}
  \centering
  \begin{tabular}{ccc} 
    \hline\hline
    The PDE model \cref{eq:rs2} & The coupled FCGL \cref{eq:cprs} 
    & Physical meaning \\[0.5ex]
    \hline \\[0ex]
    $\mu = \alpha - \gamma + \epsilon^2 \rho 
    = \frac12 \alpha + \epsilon^2 \rho$ 
    & $\rho = \dfrac{\mu-\alpha+\gamma}{\epsilon^2}$ 
    & $\rho = \text{damping ($\rho<0$)}$ \\[2ex]
    $\gamma = \frac12 \alpha$ &   \\[2ex] 
    $\delta = \frac12 \beta + \frac14 \epsilon v_g$ 
    & $v_g = \dfrac{-2\beta+4\delta}{\epsilon}$ 
    & $v_g = \text{group velocity}$ \\[2ex] 
    $\omega = 1 + \frac12 \beta - \frac14 \epsilon v_g + \epsilon^2 \nu$
    & $\nu = \dfrac{\omega-1-\beta+\delta}{\epsilon^2}$ 
    & $\nu = \text{detuning}$ \\[2ex]
    $F=4\epsilon^2\Gamma$ & $\Gamma=\dfrac{F}{4\epsilon^2}$ 
    & \begin{tabular}{@{}l@{}l@{}}
      $\Gamma = {}$ & strength of \\
      & parametric forcing
    \end{tabular} \\[2ex] 
    \hline
  \end{tabular}
\end{table}

We relate the parameters in the PDE model with the parameters in the
amplitude equations in a way that we can connect examples of localized
oscillons in both equations.  In \cref{tb:table1} all PDE
parameters are defined in terms of parameters that will appear in the
coupled FCGL equations, and vice versa.

\subsection{Linear theory}
\label{lineartheory_chap3}

With the parameters as in \cref{tb:table1}, the linear theory of
the PDE \cref{eq:rs2} at leading order is given by
\begin{equation}
  \label{eq:rslin2}
  U_{t} = 
  \left( \frac{\alpha}{2} + i \left(\frac{\beta}{2} + 1 \right) \right) U 
  + \left( \alpha+ i \beta \right) U_{xx}
  + \left( \frac{\alpha}{2} + i \frac{\beta}{2} \right) U_{xxxx},
\end{equation}
which defines a linear operator $L$ as 
\[
L U = \left( -\frac{\partial}{\partial t} + i \right) U 
+\left( \frac{\alpha}{2} + i \frac{\beta}{2} \right) 
\left( 1 + \frac{\partial^2}{\partial x^2} \right)^2 U.
\]
This is essentially the linear part of the complex Swift--Hohenberg
equation~\cite{AT2}, which has appeared in the context of nonlinear
optics~\cite{LMN} and Taylor--Couette flows~\cite{BA}.  To find all
solutions, we substitute $U=e^{\sigma t+ikx}$ into the above equation
to get the dispersion relation
\[
\sigma = i + \left( \frac{\alpha}{2} + i \frac{\beta}{2} \right)
\left( 1 - k^2 \right)^2.
\]
We assume that our problem has periodic boundary conditions, which
implies that $k \in \mathbb{R}$.  Furthermore, we require $\sigma_r = 0$
since we are considering neutral modes. The real and imaginary parts
of this equation give
\[
k = \pm 1 \quad \text{and} \quad \sigma = i.
\]
Therefore, $LU=0$, equivalent to~\cref{eq:rslin2}, implies that
neutral modes are a linear combinations of $U(x,t) = e^{i(t+x)}$ and
$U(x,t) = e^{i(t-x)}$.  Note that our choice of dispersion relation
leads to positive frequency solutions. This is not a severe
restriction, as discussed in \cref{sec:discussion}.

\subsection{Weakly nonlinear theory}

In order to apply the standard weakly nonlinear theory, we need the
adjoint linear operator $L^\dagger$. Therefore, we define an inner
product between two functions $f(x,t)$ and $g(x,t)$ by
\begin{equation}
  \label{eq:inproduct}
  \big\langle f(x,t), g(x,t) \big\rangle
  = \frac{1}{4\pi^2} \int_0^{2\pi} \int_0^{2\pi} \bar{f}(x,t) g(x,t) \,dt \,dx,
\end{equation}
where $\bar{f}$ is the complex conjugate of $f$. The adjoint linear
operator $L^\dagger$ is defined by the relation
\[
\big\langle f(x,t), L g(x,t) \big\rangle 
=\big\langle L^{\dagger} f(x,t), g(x,t) \big\rangle \qquad
\text{for all $f$ and $g$},
\]
and so, using integration by parts, 
\[
L^\dagger f = \left( \frac{\partial}{\partial t} - i 
  + \left(\frac{\alpha}{2} - i \frac{\beta}{2}\right)
  \left( 1 + \frac{\partial^2}{\partial x^2} \right)^2 \right) f.
\]
Taking the adjoint changes the sign of the
$\frac{\partial}{\partial t}$ term and takes the complex conjugate of
other terms of $L$.  The adjoint eigenfunctions are then given by
solving $L^{\dagger}f=0$; the solutions are also linear combinations
of $e^{i(t\pm x)}.$

We expand $U$ in powers of the small parameter $\epsilon$:
\begin{equation}\label{eq:uexpand1}
  U = \epsilon U_1 + \epsilon^2 U_2  + \epsilon^3 U_3 + \cdots,
\end{equation}
where $U_1$, $U_2$, $U_3$, \dots are $O(1)$ complex functions.  We
will derive solutions $U_1$, $U_2$, $U_3$, \dots at each order of
$\epsilon$.

At $O(\epsilon)$, the linear theory arises and we find $LU_1 = 0$. The
solution $U_1$ takes the form
\begin{equation}\label{eq:solution_U1}
U_1=A(X,T)e^{i(t +x)}+B(X,T)e^{i(t -x)},
\end{equation}
where $A$ and $B$ represent the amplitudes of the left and right
travelling waves. They are functions of $X$ and $T$, the long and slow
scale modulations of space and time variables:
$$T = \epsilon^2 t,\quad{}\text{and}\quad{} X = \epsilon x.$$ The
multiple scale expansion below will determine the evolution equations
for $A(X,T)$ and $B(X,T)$.

At second order in $\epsilon$, we get $L U_2 = 0$: the
$\frac{\partial^2 U_1}{\partial x \partial X }$ term cancels with the
$\frac{\partial^4 U_1}{\partial x^3 \partial X }$ term.  We would have
had a forcing term at this order if we had not ensured that the group
velocity is $O(\epsilon)$. The equation at this order is solved by
setting $U_2 = 0$.

At third order in $\epsilon$, we get
\begin{equation}\label{eq:ampAB1}
  \begin{aligned}
    \frac{\partial U_1}{\partial T}
    &= L U_3 + (\rho+i\nu) U_1 
    + (\alpha+i\beta) \frac{\partial^2 U_1}{\partial X^2}
    + 3 (\alpha + i \beta) \frac{\partial^4U_1}{\partial x^2\partial X^2} \\
    &\quad{} + i v_g \frac{\partial^4 U_1}{\partial x^3 \partial X}
    + 4i \Gamma \cos(2t) \Re(U_1) + C|U_1|^2 U_1.
  \end{aligned}
\end{equation}
The linear operator $L$ is singular so we must apply a solvability
condition: we take the inner product between the adjoint eigenfunction
$e^{i(t+x)}$ and equation \cref{eq:ampAB1}, which gives
\begin{equation}\label{eq:rs4}
\begin{aligned}
  \left\langle e^{i(t+x)} ,\frac{\partial U_1}{\partial T} \right\rangle
  &= \left\langle e^{i(t+x)}, LU_3 \right\rangle 
  + (\rho + i \nu) \left\langle e^{i(t+x)}, U_1 \right\rangle 
  + (\alpha + i \beta) \left\langle e^{i(t+x)}, 
    \frac{\partial^2 U_1}{\partial X^2} \right\rangle \\
  &\quad{} + 3 (\alpha + i \beta) \left\langle e^{i(t+x)}, 
    \frac{\partial^4U_1}{\partial x^2\partial X^2} \right\rangle
  + i v_g \left\langle e^{i(t+x)}, 
    \frac{\partial^4 U_1}{\partial x^3 \partial X} \right\rangle \\
  &\quad{} + 4i \Gamma \left\langle e^{i(t+x)}, \cos(2t) \Re(U_1) \right\rangle 
  + C \left\langle e^{i(t+x)}, |U_1|^2 U_1 \right\rangle.
\end{aligned}
\end{equation}
We have
$\langle e^{i(t+x)}, L U_3 \rangle = \langle L^\dagger e^{i(t+x)}, U_3
\rangle = 0$,
so $U_3$ is removed and the above equation becomes an equation in
$U_1$ only. Substituting the solution $U_1$ leads to
\begin{equation}\label{eq:rs5}
\begin{aligned}
  &\left\langle e^{i(t+x)}, \frac{\partial}{\partial T} 
    (A e^{i(t+x)} + B e^{i(t-x)}) \right\rangle \\
  &= (\rho+i\nu) \left\langle e^{i(t+x)}, 
    A e^{i(t+x)} + B e^{i(t-x)} \right\rangle \\
  &\qquad{} + (\alpha + i \beta) \left\langle e^{i(t+x)},
    \frac{\partial^2}{\partial X^2} (A e^{i(t+x)} + B e^{i(t-x)})
  \right\rangle \\
  &\qquad{} + 3 (\alpha + i \beta) \left\langle e^{i(t+x)},
    \frac{\partial^4}{\partial x^2\partial X^2} (A e^{i(t+x)} + B e^{i(t-x)})
  \right\rangle \\
  &\qquad{} + i v_g \left\langle e^{i(t+x)},
    \frac{\partial^4}{\partial x^3 \partial X} (A e^{i(t+x)} + B e^{i(t-x)})
  \right\rangle \\
  &\qquad{} + 4i \Gamma \left\langle e^{i(t+x)},
    \tfrac{1}{2} \cos(2t) (A e^{i(x+t)} + B e^{i(t-x)} + \bar{A} e^{-i(t+x)}
    + \bar{B}e^{-i(t-x)}) \right\rangle \\
  &\qquad{} + C \left\langle e^{i(t+x)}, 
    (|A|^2 + A \bar{B} e^{2ix} + \bar{A} B e^{-2ix} + |B|^2) 
    (A e^{i(t+x)} + B e^{i(t-x)}) \right\rangle.
\end{aligned}
\end{equation}
After we compute the left and right hand sides of the above equation
term by term, we get equations for the amplitudes $A(X,T)$ and
$B(X,T)$:
\begin{equation}\label{eq:cprs}
\begin{aligned}
  \frac{\partial A}{\partial T}
  &= (\rho + i \nu) A - 2 (\alpha + i \beta) \frac{\partial^2 A}{\partial X^2}
  + v_g \frac{\partial A}{\partial X} + C (|A|^2 + 2 |B|^2) A
  +i \Gamma \bar{B}, \\
  \frac{\partial B}{\partial T}
  &= (\rho + i \nu) B - 2 (\alpha + i \beta) \frac{\partial^2 B}{\partial X^2}
  - v_g \frac{\partial B}{\partial X} + C (2 |A|^2 + |B|^2) B 
  + i \Gamma \bar{A}.
\end{aligned}
\end{equation}
Thus the PDE model has been reduced to the coupled FCGL equations in
the weak damping, weak detuning, small group velocity and small
amplitude limit.  In equations~\cref{eq:cprs} the group velocity terms
have different signs, which makes the envelopes travel in opposite
directions. The $-2\alpha\frac{\partial^2A}{\partial X^2}$ may make
the above equations look like they are ill posed, but recall that
$\alpha<0$.

\section{Properties of the coupled  FCGL  equations}

Following \cite{R} we can identify the symmetries and how they affect
the structure of \cref{eq:cprs}. The original system is invariant
under translations in~$x$: Replacing $x$ by $x+\phi^*$, where $\phi^*$
is arbitrary, we get
\[
U(x+\phi^*, t) = A(X+\epsilon\phi^*,T) \, e^{i(t+x+\phi^*)} 
+ B(X+\epsilon\phi^*,T) \, e^{i(t-x-\phi^*)},
\]
which is also a solution of the problem. This translation has the
effect of shifting $X$ to $X + \epsilon\phi^*$, and changing the phase
of $A$ and~$B$: If we suppress the change from $X$ to
$X+\epsilon\phi^*$, then \cref{eq:cprs} is equivariant under
\[
A \rightarrow A e^{i\phi^*}, \quad B \rightarrow B e^{-i\phi^*},
\]
which is therefore a symmetry of~\cref{eq:cprs}.
Equations~\cref{eq:cprs} are also invariant under translations in~$X$,
but this is an artifact of the truncation at cubic order~\cite{Me}.
Similarly, we can reflect in~$x$, which leads to the symmetry
$A \leftrightarrow B$, $\partial_x \leftrightarrow -\partial_x$.

Amplitude equations associated with a Hopf bifurcation (a weakly
damped Hopf bifurcation in this case) usually have time translation
symmetry, which manifests as equivariance under phase shifts of the
amplitudes. However, the underlying PDE is non-autonomous, and so
rotating $A$ and $B$ by a common phase is not a symmetry of
\cref{eq:cprs}.  Equations~\cref{eq:cprs} do posess $T$-translation
symmetry, but this is also an artifact.

The parametric forcing provides an interesting coupling between the
left and right travelling waves with amplitudes $A$ and $B$, which
means that solutions or symmetries that one might expect at first
glance, are in fact not present. For example, the coupling terms in
the coupled FCGL equations make it impossible to find pure travelling
waves; i.e., $A\neq 0$, $B=0$ is not a solution of~\cref{eq:cprs}.
Also, solutions with $A = B$ exist only if $v_g$ is zero, which
generically it is not. Finally, steady standing wave solutions (which
are typically seen in Faraday wave experiments) have $B(X) = A(-X)$;
substituting this into~\cref{eq:cprs} yields a nonlocal equation that
is not a PDE, though all solutions we present in this paper are in
this category.

\subsection{The zero solution}
\label{section331}

The stability of the zero state under small perturbations
with complex growth rate $s$ and real wavenumber $q$ can be studied by
linearizing~\cref{eq:cprs}, writing $A$ and~$B$ as
\[
A = \hat{A} e^{sT+iqX}, \quad \text{and} \quad
B = \hat{B} e^{\bar{s}T-iqX},
\]
where $|\hat{A}| \ll 1$, $|\hat{B}| \ll 1$ and
$\hat{A}, \hat{B} \in \mathbb{C}$.  We choose
$\hat{B} e^{\bar{s}T-iqX}$ in order that the exponential term will
cancel in the next step.  Substituting this into equation
\cref{eq:cprs}, linearizing and taking the complex conjugate of the
second equation gives:
\begin{equation}
  \label{eq:steady}
  \begin{aligned}
    s\hat{A} &= (\rho+i\nu) \hat{A} + 2 (\alpha+i\beta) q^2 \hat{A} 
    + i v_g q \hat{A} + i \Gamma \bar{\hat{B}}.\\
    s\bar{\hat{B}} &= (\rho-i\nu) \bar{\hat{B}} 
    + 2 (\alpha-i\beta) q^2 \bar{\hat{B}} - i v_g q \bar{\hat{B}} 
    - i \Gamma \hat{A}.
  \end{aligned}
\end{equation}
This is a linear homogeneous system of equations, so there is a
nontrivial solution only when its determinant is zero. The imaginary
part of the determinant equals $2 s_i (\rho + 2 \alpha q^2 - s_r)$,
where $s_r$ and $s_i$ denote the real and imaginary part of~$s$. We
are interested in locating the bifurcation where zero solution is
neutrally stable, so $s_r = 0$. Since $\rho$ and $\alpha$ are
negative, the determinant can only be zero if $s_i = 0$. Thus, there
is no Hopf bifurcation, and the neutral stability condition is
$s=0$. Setting the real part of the determinant of~\cref{eq:steady}
equal to zero leads to:
\begin{equation}
  \label{eq:q_Gamma}
  (\rho + 2 \alpha q^2)^2 + (\nu + 2 \beta q^2 + v_g q)^2 = \Gamma^2.
\end{equation} 
The stability of the zero state changes when $\Gamma=\Gamma_c$, the
minimum of the neutral stability curve, and the non-zero flat state is
created with $q=q_c$. This corresponds to a uniform pattern in the PDE
\cref{eq:rs2} with wavenumber $k_c = 1 + \epsilon q_c$. The critical
wavenumber $q_c$ can be computed by minimizing the left-hand side of
equation~\cref{eq:q_Gamma}. Differentiating with respect to~$q$
yields the following cubic equation in~$q$:
\begin{equation}
  \label{eq:q_Gamma_min}
  4 \alpha q (\rho + 2 \alpha q^2) 
  + (4 \beta q + v_g) (\nu + 2\beta q^2 + v_g q) = 0.
\end{equation}
Solving this gives $q_c$, the critical wavenumber, which is positive
if $\nu v_g<0$ and negative if $\nu v_g > 0$. Substituting $q = q_c$
into~\cref{eq:q_Gamma} gives~$\Gamma_c$.

\subsection{Standing waves}
\label{section333}

Now we look at steady equal-amplitude states of the form
$A = R_0 e^{i(qX+\phi_1)}$ and $B = R_0 e^{i(-qX+\phi_2)}$, where
$R_0$ and $q$ are real, and $\phi_1$ and~$\phi_2$ are the
phases. These represent uniform standing wave patterns with wavenumber
$1+\epsilon q$ in $U(x)$.  We substitute this into
equations~\cref{eq:cprs}, which yields, assuming that $R_0$ is not
zero,
\[
0 = (\rho + i \nu) + 2 (\alpha + i \beta) q^2 + i v_g q + 3 C R_0^2 
+ i \Gamma e^{-i\Phi},
\]
where $\Phi = \phi_1 + \phi_2$. This is the same equation obtained for
steady constant-amplitude solutions of the single FCGL
equation~\cref{eq:fcgl}, but with a group velocity term.  The real
and imaginary parts of the above equation are
\begin{equation}
  \label{eq:re2}
  \begin{aligned}
    \text{Re:} \quad 
    0 &= \rho + 2 \alpha q^2 + 3 C_r R_0 ^2 + \Gamma \sin\Phi.\\
    \text{Im:} \quad 
    0 &= \nu + 2 \beta q^2 + v_g q + 3C_i R_0^2 + \Gamma \cos\Phi.
  \end{aligned}
\end{equation}
We eliminate $\Phi$ by using the identity
$\cos^2\Phi + \sin^2\Phi = 1$ to give the following polynomial
equation for~$R_0$:
\begin{multline}
  \label{eq:poly3}
  0 = 9 (C_r^2 + C_i^2) R_0^4 
  + 6 \left( (\rho + 2 \alpha q^2) C_r 
    + (\nu + v_g q + 2 \beta q^2) C_i \right) R_0^2 \\
  + (\rho + 2\alpha q^2)^2 + (\nu + v_g q + 2\beta q^2)^2
  - \Gamma^2.
\end{multline}
This is a quadratic equation in $R_0^2$ and its discriminant is given
by
\begin{multline*}
\Delta = 36 \left( (\rho+2\alpha q^2) C_r + (\nu+v_g q+2\beta q^2) C_i \right)^2
\\
- 36 \left( (\rho+2\alpha q^2)^2 + (\nu+v_g q+2\beta q^2)^2 - \Gamma^2 \right)
(C_r ^2+C_i ^2).
\end{multline*}

\begin{figure}
  \centering
  \includegraphics[width=0.5\textwidth]{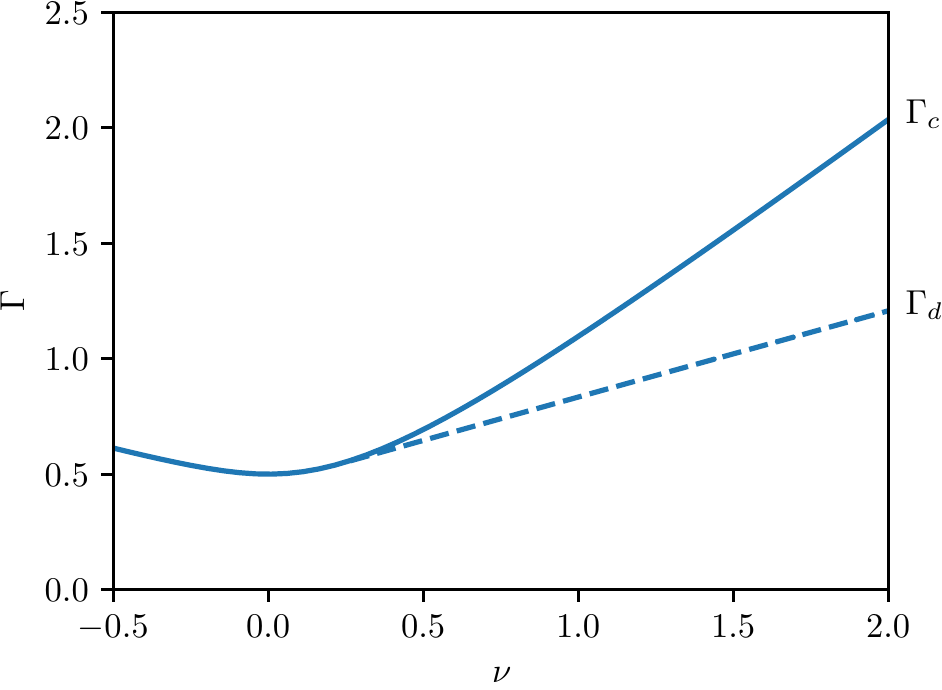}
  \caption{The $(\nu, \Gamma)$ parameter plane of the coupled FCGL
    equations \cref{eq:cprs} with $\rho=-0.5$, $\alpha=-0.5$,
    $\beta=1$, $v_g=-0.5$ and $C = -1-2.5i$. These parameters with
    $\epsilon = 0.1$ correspond to the prameters of the model
    PDE~\cref{eq:rs2} used in the figures in
    \cref{sec:numerics}. The solid line shows the primary
    pitchfork bifurcation at $\Gamma_c$, where the zero state becomes
    unstable to perturbations with wavenumber~$q_c$.  The dash line
    shows the saddle-node bifurcation at $\Gamma_d$.}
  \label{fig:nu-gamma_cprs}
\end{figure}

Examination of the polynomial \cref{eq:poly3} shows that when the
forcing amplitude $\Gamma$ reaches
$((\rho + 2 \alpha q^2)^2 + (\nu + v_g q + 2 \beta q^2)^2)^{1/2}$, a
subcritical bifurcation occurs provided that
$(\rho + 2 \alpha q^2) C_r + (\nu + v_g q + 2 \beta q^2) C_i<0$.
Spatially oscillatory states $A^-_{sp}$ and $B^-_{sp}$ are created,
which turn into $A^+_{sp} $ and $B^+_{sp}$ states at a saddle-node
($\Delta=0$) bifurcation at $\Gamma=\Gamma_d$, with
\begin{equation}
  \label{eq:saddle_node_curve}
  \Gamma_d = \sqrt{(\rho + 2 \alpha q^2)^2
    + (\nu + v_g q + 2 \beta q^2)^2
    - \frac{((\rho + 2 \alpha q^2) C_r + (\nu + v_g q + 2 \beta q^2) C_i)^2}
    {C_r^2 + C_i^2}}.
\end{equation}
\Cref{fig:nu-gamma_cprs} shows equations \cref{eq:q_Gamma} and
\cref{eq:saddle_node_curve} in the $(\nu, \Gamma)$ parameter plane
where we have taken $q = q_c$ from~\cref{eq:q_Gamma_min}. The values
of the parameters $\rho$, $\alpha$, $\beta$, $v_g$, $C_r$ and $C_i$ in
the figure correspond to the parameters in the figures in
\cref{sec:numerics} with $\epsilon = 0.1$. The primary bifurcation
changes from supercritical to subcritical when
$(\rho + 2 \alpha q^2) C_r + (\nu + v_g q + 2 \beta q^2) C_i = 0$,
which is at $\nu=0.2228$ for the parameter values in
\cref{fig:nu-gamma_cprs}.  Localized solutions can be found in the
bistability region between $\Gamma_c$ and $\Gamma_d$.

\subsection{Localized solutions}

In order to find localized solutions of the coupled FCGL
equations~\cref{eq:cprs}, one might attempt an ansatz of the form 
\[
A = R_0(X,T) \, e^{i(qX+\phi_1)} \quad\text{and}\quad 
B = \bar{R}_0(X,T) \, e^{i(-qX+\phi_2)} 
\]
with $R_0$ complex and $q, \phi_1, \phi_2$ real. This is a spatially
modulated version of the standing wave studied in the previous
section. However, the coupled FCGL equations admit no solution of this
form, even if $v_g = 0$. Other standing wave ansatzes are possible,
e.g.  $A = R_0(X) \, e^{i(qX+\phi_1)}$ and
$B = \bar{R}_0(-X) \, e^{i(-qX+\phi_2)}$, but we have not explored
these further.

We were able to find analytic expressions for localized solutions of
the coupled FCGL equations by taking further asymptotic limits (see
\cref{section34}).  To motivate the subsequent calculations, we
present some numerical examples of stable spatially localized
oscillons in the coupled FCGL equations found by using the same
numerical method as in \cref{sec:numerics} on a periodic domain of
size $20\pi$.  We take the same parameter values as before:
$\rho=-0.5,$ $\nu=2$, $\alpha=-0.5$, $\beta=1$, and $C=-1-2.5i$.

\begin{figure}
  \centering
  \includegraphics[width=\textwidth]{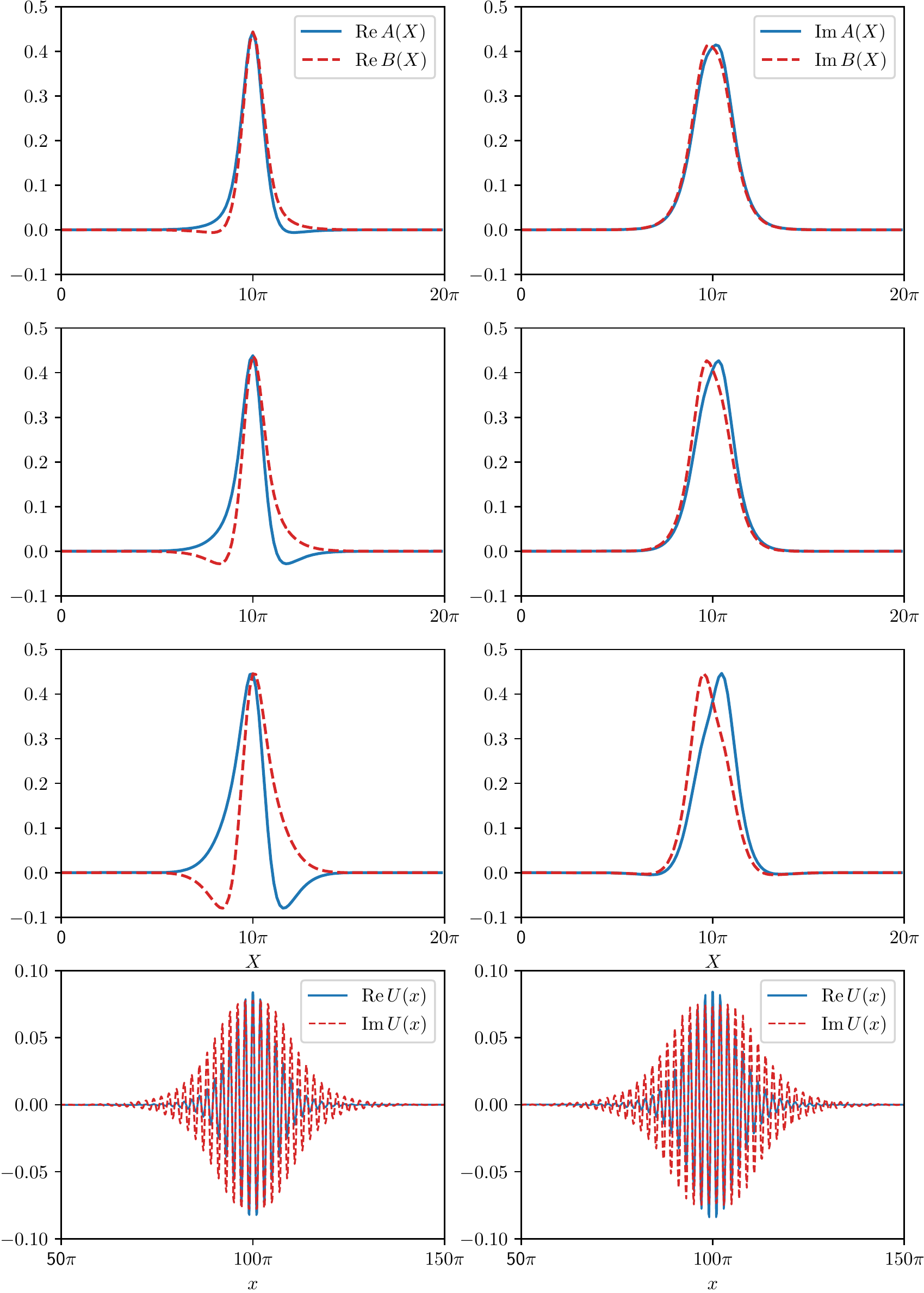} 
  \caption{Stationary solutions to the coupled FCGL equations
    \cref{eq:cprs} with $\rho = -0.5$, $\nu = 2$, $\alpha = -0.5$,
    $\beta = 1$ and $C = -1 - 2.5i$. Top row: $v_g = -0.2$ and $\Gamma
    = 1.46$.  Second row: $v_g = -0.5$ and $\Gamma = 1.45$. Third row:
    $v_g = -1$ and $\Gamma = 1.43$. Bottom row: Approximate
    solutions~$U(x)$ of the PDE model~\cref{eq:rs2} reconstructed from
    the solutions $A(X), B(X)$ to the coupled FCGL equations assuming
    $\epsilon = 0.1$; the left and right plots correspond to the top
    and third rows, respectively.}
  \label{fig:cpr_ex1}
\end{figure}

The top row of \cref{fig:cpr_ex1} shows an example of a localized
oscillon in the coupled FCGL equations with $v_g = -0.2$. As we
increase the magnitude of the group velocity~$v_g$ to $v_g = -0.5$
(second row) and $v_g = -1$ (third row) and change the forcing
strength~$\Gamma$ so that we are still in the region where the
localized solution is stable, we can see that $A$ and $B$ start to
move apart, pulled in opposite directions by the group velocity
term. We can use these solutions to the coupled FCGL equations to
reconstruct first-order approximations to solutions of the PDE
model~\cref{eq:rs2} with the help
of~\cref{eq:uexpand1,eq:solution_U1}; this is shown in the bottom row
of \cref{fig:cpr_ex1}.

\begin{figure}
  \centering
  \includegraphics[width=\linewidth]{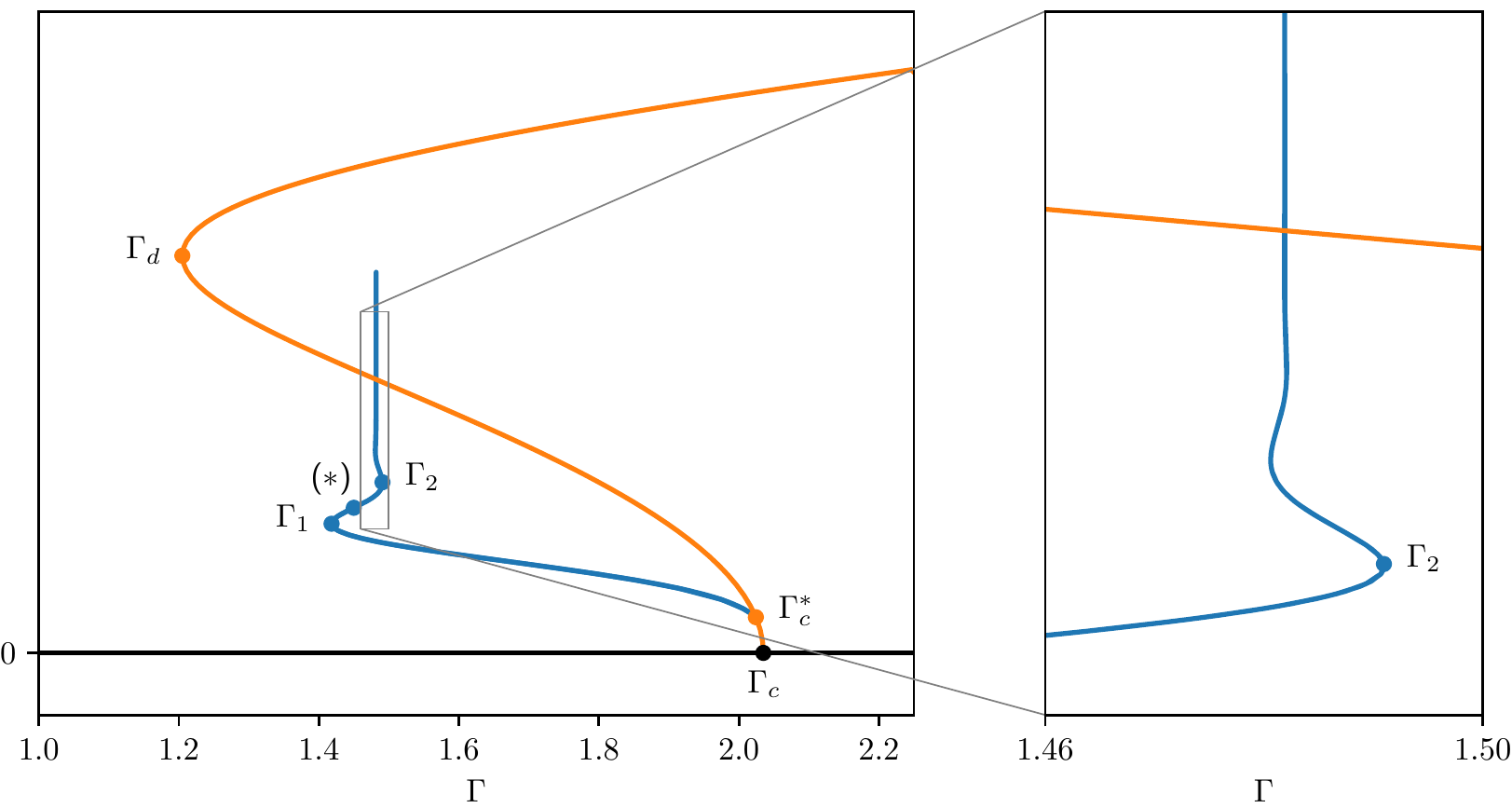}
  \caption{Bifurcation diagram of the coupled FCGL equations with
    parameters $\rho = -0.5$, $\nu = 2$, $\alpha = -0.5$, $\beta = 1$,
    $C = -1-2.5i$ and $v_g = -0.5$ (corresponding to
    \cref{fig:bif_pde}). The bifurcations are at $\Gamma_c =
    2.035$, $\Gamma_d = 1.206$, $\Gamma_c^* = 2.024$, $\Gamma_1 =
    1.418$ and $\Gamma_2 = 1.491$. The solution marked~($*$) is shown
    in the middle row of \cref{fig:cpr_ex1}. Numerical results
    with AUTO suggest that snaking continues beyond~$\Gamma_2$, but
    it is too small to see.}
  \label{fig:cfcgl_bifdiag}
\end{figure}

We also computed the bifurcation diagram of the coupled FCGL equations
on a domain of size~$20\pi$ using AUTO~\cite{Doe}. The critical
wavenumber with the above parameter values is
$q_c = 0.09950 \approx \frac{1}{10}$, so the periodic solution fits
almost perfectly in this domain.  The result is shown in
\cref{fig:cfcgl_bifdiag}. The branch of periodic solution bifurcates
from the zero solution at $\Gamma_c = 2.035$ and has a fold at
$\Gamma_d = 1.206$. Using the relation $F = 4 \epsilon^2 \Gamma$, we
can compute the corresponding values of forcing in the model
PDE~\cref{eq:rs2} as $0.08140$ and $0.04820$, which agree well with
the values of $F_c = 0.08173$ and $F_d = 0.04811$ found in
\cref{fig:bif_pde} when we applied AUTO directly to the model PDE.

Going back to the coupled FCGL equations, we see a secondary
bifurcation at $\Gamma = 2.024$ where a branch of localized solutions
bifurcates from the branch of periodic solutions.
The localized branch has folds at $\Gamma_1 = 1.418$ and $\Gamma_2 =
1.491$. The corresponding $F$~values in terms of the parameters of the
model PDE are $0.05673$ and $0.05964$, which again agree well with the
values of $F_1 = 0.05666$ and $F_2 = 0.05948$ found in
\cref{fig:bif_pde}.

As shown in \cref{fig:cfcgl_bifdiag}, the localized branch in the
bifurcation diagram of the coupled FCGL equations continues to snake
upwards after~$\Gamma_2$. We believe that these exhibit collapsed
snaking, where the saddle node points asymptote to one value
of~$\Gamma$ as one goes up the branch~\cite{MBK}. However, the
bifurcation diagram shows that the branch of localized solutions
suddenly stops. In fact, AUTO turns around at that point. We believe
that this may be caused by AUTO having difficulty handling the phase
symmetry in the coupled FCGL equations, and that in reality the branch
of localized solutions joins with the branch of periodic solutions
near the fold at~$\Gamma_d$, as it does in the bifurcation diagram of
the model PDE in \cref{fig:bif_pde}.

\section{Reduction to the real Ginzburg--Landau equation}
\label{section34}

In this section we will reduce the coupled FCGL equations to the real
Ginzburg--Landau equation close to the subcritical bifurcation from
the zero solution to the constant amplitude state. The reduction was
done by Riecke \cite{RI} in the supercritical case.

We take the complex conjugate of the second equation of
\cref{eq:cprs}, so the coupled FCGL equations become
\begin{equation}
  \label{eq:cprs_reduction}
  \begin{aligned}
    \frac{\partial A}{\partial T}
    &= D_1 A + D_2 \frac{\partial^2 A}{\partial X^2}
    + v_g \frac{\partial A}{\partial X} 
    + C (|A|^2 + 2 |B|^2) A + i \Gamma \bar{B}, \\
    \frac{\partial\bar{B}}{\partial T}
    &= \bar{D}_1 \bar{B} + \bar{D}_2 \frac{\partial\bar{B}}{\partial X^2}
    - v_g \frac{\partial\bar{B}}{\partial X}
    + \bar{C} (2 |A|^2 + |B|^2) \bar{B} - i \Gamma A.
  \end{aligned}
\end{equation}
For simplicity, we write 
\begin{equation}
  \label{eq:D1_D2}
  D_1 = \rho + i \nu \quad\text{and}\quad
  D_2 = -2 (\alpha + i \beta). 
\end{equation}
In order to reduce the coupled FCGL equation to the real
Ginzburg--Landau equation, we apply weakly nonlinear theory close to
onset, writing
\[
\Gamma = \Gamma_c (1 + \epsilon_2^2 \Gamma_2),
\]
where $0<\epsilon_2\ll 1$, and $\Gamma_c$ is the critical forcing at
critical wavenumber $q_c$, and $ \Gamma_2$ is the new bifurcation
parameter. We expand the solution in powers of the new small parameter
$\epsilon_2$ as follows 
\[
\begin{bmatrix} A \\\bar{B} \end{bmatrix}
=
\begin{bmatrix}
  \epsilon_2 A_1 + \epsilon_2^2 A_2 + \epsilon_2^3 A_3 + \cdots \\
  \epsilon_2 \bar{B}_1 + \epsilon_2^2 \bar{B}_2 + \epsilon_2^3 \bar{B}_3
  + \cdots
\end{bmatrix}.
\]
From \cref{section331}, the growth rate is real with frequency
zero (locked to the forcing), so we scale
\[
\frac{\partial}{\partial T} \rightarrow 
\epsilon_2^2 \frac{\partial}{\partial\tilde{T}},
\]
and the preferred wavenumber $q_c \neq 0$, so   
\[
\frac{\partial}{\partial X} \rightarrow \frac{\partial}{\partial X}
+ \epsilon_2 \frac{\partial}{\partial\tilde{X}},
\]
where $\tilde{X}$ and $\tilde{T}$ are very long space and slow time
scales.

At $O(\epsilon_2)$,  we have
\begin{align*}
  0 &= D_1 A_1 + D_2 \frac{\partial^2 A_1}{\partial X^2}
      + v_g \frac{\partial A_1}{\partial X} + i \Gamma_c \bar{B}_1,\\
  0 & = \bar{D}_1 \bar{B}_1 + \bar{D}_2 \frac{\partial\bar{B}_1}{\partial X^2}
      - v_g \frac{\partial\bar{B}_1}{\partial X} - i \Gamma_c A_1.
\end{align*}
We can solve the above system by assuming that
\begin{equation}
  \label{eq:A1_B1}
  A_1 = P(\tilde{X}, \tilde{T}) e^{iq_c X} 
  \quad \text{and} \quad
  B_1 = Q(\tilde{X}, \tilde{T}) e^{-iq_c X}.
\end{equation}
At this order of $\epsilon_2$, the coupled FCGL
equations become
\begin{equation}
  \label{eq:cprs_reduction_linear}
  \begin{aligned}
    0 &= D_1 P - D_2 q^2_c P + i v_g q_c P + i \Gamma_c \bar{Q}, \\
    0 &= \bar{D}_1 \bar{Q} - \bar{D}_2 q^2_c \bar{Q} - i v_g q_c\bar{Q} 
    - i \Gamma_c P.
\end{aligned}
\end{equation}
These can be solved as in \cref{section331}. 

Additionally, from the first equation of
\cref{eq:cprs_reduction_linear} we get a phase relation between $P$
and~$Q$:
\begin{equation}
  \label{eq:relationship2}
  \bar{Q} = P e^{i\phi} \quad\text{where}\quad
  e^{i\phi} = -\frac{D_1 + i v_g q_c - D_2 q^2_c}{i \Gamma_c}.
\end{equation}
The fraction in the above equation has modulus~1, so the phase~$\phi$
is real.

At $O(\epsilon_2^2)$, equations \cref{eq:cprs_reduction} become
\begin{equation}
  \label{eq:order2_eqs}
  \begin{aligned}
    0 & = D_1 A_2
    + D_2 \frac{\partial^2 A_2}{\partial X^2} 
    + v_g \frac{\partial A_2}{\partial X} + i \Gamma_c \bar{B}_2 
    + v_g \frac{\partial A}{\partial\tilde{X}} e^{iq_c X}
    + 2i D_2 q_c \frac{\partial A}{\partial\tilde{X}} e^{iq_c X}, \\
    0 &= \bar{D}_1 \bar{B}_2 
    + \bar{D}_2 \frac{\partial^2 \bar{B}_2}{\partial X^2} 
    - v_g \frac{\partial\bar{B}_2}{\partial X} - i \Gamma_c A_2 
    - v_g \frac{\partial\bar{B}}{\partial\tilde{X}} e^{iq_c X}
    + 2i \bar{D}_2 q_c \frac{\partial\bar{B}}{\partial\tilde{X}} e^{iq_c X}.
  \end{aligned}
\end{equation}
At this stage we would normally define a linear operator in order to
impose a solvability condition. In this case, the solvability
condition can be deduced directly by setting
\begin{equation}
  \label{eq:A2mode}
  A_2 = P_2 e^{i q_c X} + \cdots
  \quad \text{and} \quad
  \bar{B}_2 = \bar{Q}_2 e^{i q_c X} + \cdots
\end{equation}
where the dots stand for the other Fourier components. This focuses
the attention on the $e^{i q_c X} $ component
of~\cref{eq:order2_eqs}, which is the only component to have an
inhomogeneous part and for which the linear operator is singular.
Substituting these expressions for~$A_2$ and $\bar{B}_2$
into~\cref{eq:order2_eqs} and using~\cref{eq:relationship2} leads to
the following:
\begin{equation}
  \label{eq:matrix2}
  \begin{bmatrix}
    D_1 + i v_g q_c - D_2 q_c ^2 & i \Gamma_c \\
    -i \Gamma_c & \bar{D}_1 - i v_g q_c -\bar{D}_2 q_c^2
  \end{bmatrix}
  \begin{bmatrix} P_2 \\ \bar{Q}_2 \end{bmatrix}
  +
  \begin{bmatrix}
    v_g + 2 i q_c D_2 \\
    (-v_g + 2 i q_c \bar{D}_2) e^{i\phi}
  \end{bmatrix}
  \frac{\partial P}{\partial\tilde{X}}
  = \begin{bmatrix} 0 \\ 0 \end{bmatrix},
\end{equation}
where $e^{i\phi}$ is defined in~\cref{eq:relationship2}. The square
matrix is singular since it is the same one that appears in the linear
theory; see~\cref{eq:steady}. We multiply the first line by
$i\Gamma_c$ and the second line by $D_1 + iv_g q_c - D_2 q_c^2$, which
is effectively the left eigenvector of the matrix, then add both lines
and use \cref{eq:q_Gamma} and~\cref{eq:relationship2}, ending up with
\[
\left( i \Gamma_c (v_g + 2 i q_c D_2) + 
  \frac{(v_g - 2 i q_c \bar{D}_2)(D_1 + i v_g q_c - D_2 q_c ^2)^2}{i\Gamma_c} 
\right) \frac{\partial P}{\partial\tilde{X}} = 0.
\]
Since $\frac{\partial P}{\partial\tilde{X}} \neq 0$, we need
\[
- \Gamma_c^2 (v_g + 2 i q_c D_2) + (v_g - 2 i q_c \bar{D}_2)
(D_1 + i v_g q_c - D_2 q_c ^2)^2 = 0
\]
After substituting~\cref{eq:q_Gamma}, we find that this is the same
as \cref{eq:q_Gamma_min}, which is satisfied since $q_c$ is at the
minimum of the neutral stability curve.

From the top line of \cref{eq:matrix2}, we have the solution
\[
\bar{Q}_2 = - \left( 
  \frac{v_g + 2 i q_c D_2}{i \Gamma_c} \frac{\partial A}{\partial\tilde{X}} 
  + \frac{D_1 + i v_g q_c - D_2 q_c ^2}{i \Gamma_c} A_2
\right).
\]
Thus, we have $P_2$ arbitrary at this order of $\epsilon_2$; we can
set $P_2 = 0$, and so, restoring the $e^{iq_c X}$~factor, we have
\begin{equation}
  \label{eq:A2_B2}
  A_2 = 0
  \quad \text{and} \quad
  \bar{B}_2 = -\frac{v_g + 2 i q_c D_2}{i \Gamma_c} 
  \frac{\partial P}{\partial\tilde{X}} e^{iq_c X}.
\end{equation}
At $O(\epsilon_2^3)$ the problem has the following structure (after
using $A_2=0$):
\begin{equation}\label{eq:nonzero_real_amp}
  \begin{aligned}
    \frac{\partial A_1}{\partial\tilde{T}}
    &= D_1 A_3 + D_2 \frac{\partial^2 A_3}{\partial X^2}
    + v_g \frac{\partial A_3}{\partial X}+ i \Gamma_c \bar{B}_3 \\
    &\qquad + D_2 \frac{\partial^2 A_1}{\partial\tilde{X}^2} 
    + i \Gamma_c \Gamma_2 \bar{B}_1 +C (|A_1|^2 + 2 |B_1|^2) A_1, \\
    \frac{\partial\bar{B}_1}{\partial\tilde{T}}
    &= \bar{D}_1 \bar{B}_3 + \bar{D}_2 \frac{\partial^2\bar{B}_3}{\partial X^2}
    - v_g \frac{\partial\bar{B}_3}{\partial X} - i\Gamma_c A_3 
    + 2\bar{D}_2 \frac{\partial^2\bar{B}_2}{\partial X\partial\tilde{X}} \\
    &\qquad - v_g \frac{\partial\bar{B_2}}{\partial \tilde{X}}
    + \bar{D}_2 \frac{\partial^2\bar{B}_1}{\partial\tilde{X}^2}
    - i \Gamma_c \Gamma_2 A_1 + \bar{C} (2 |A_1|^2 + |B_1|^2) B_1.
  \end{aligned}
\end{equation}
We focus on the $e^{iq_cX}$ Fourier modes as before and write
\[
A_3 = P_3 e^{iq_c X} + \cdots
\quad \text{and} \quad
\bar{B}_3 = \bar{Q}_3 e^{iq_c X} + \cdots.
\]
As at order $\epsilon_2^2$, we multiply the first equation by
$i \Gamma_c$ and the second equation by
$D_1 + i v_g q_c - D_2 q_c ^2$, and then add them to eliminate $P_3$
and $Q_3$, finding
\begin{equation}
  \begin{aligned}
    &i \Gamma_c \frac{\partial A_1}{\partial\tilde{T}}
    + (D_1 + i v_g q_c - D_2 q_c^2) \frac{\partial\bar{B}_1}{\partial\tilde{T}}
    \\
    &\quad = i \Gamma_c D_2 \frac{\partial^2 A_1}{\partial\tilde{X}^2}
    - \Gamma_c^2 \Gamma_2 \bar{B}_1
    + i \Gamma_c C (|A_1|^2 + 2 |B_1|^2) A_1 \\
    &\qquad + 2 (D_1 + i v_g q_c - D_2 q_c^2) \bar{D}_2 
    \frac{\partial^2}{\partial X\partial \tilde{X}}\bar{B}_2 \\
    &\qquad - (D_1 + i v_g q_c - D_2 q_c^2)
    \left( v_g \frac{\partial\bar{B}_2}{\partial\tilde{X}}
      - \bar{D}_2 \frac{\partial^2\bar{B}_1}{\partial\tilde{X}^2} \right) \\
    &\qquad - i \Gamma_c \Gamma_2 (D_1 + i v_g q_c - D_2 q_c^2) A_1
    +\bar{C} (D_1 + i v_g q_c - D_2 q_c^2) (2 |A_1|^2 + |B_1|^2) B_1.
  \end{aligned}
\end{equation}
We use \cref{eq:A1_B1}, \cref{eq:relationship2} and~\cref{eq:A2_B2}
to substitute $A_1$, $B_1$ and $B_2$ into the above equation and
divide by the common factor of $e^{iq_c X}$. After some manipulation
with the help of~\cref{eq:q_Gamma} and~\cref{eq:D1_D2}, this gives
the real Ginzburg--Landau equation
\begin{multline}
  \label{eq:amplitude_real_GL}
  \frac{\partial P}{\partial\tilde{T}} =
  - \frac{\Gamma_c^2 \Gamma_2}{\rho + 2 \alpha q_c^2} P
  - \frac{4 \rho \alpha + 4 \nu \beta + v_g^2 + 12 v_g \beta q_c 
    + 24 (\alpha^2 + \beta^2) q_c^2}{2 \rho + 4 \alpha q_c^2}
  \frac{\partial^2 P}{\partial\tilde{X}^2} \\
  +3 \left( C_r + \frac{\nu + v_g q_c + 2 \beta q_c^2}{\rho + 2 \alpha q_c^2}
    C_i \right) |P|^2 P.
\end{multline}
Flat solutions of this equation correspond to the simple
constant-amplitude solutions discussed in \cref{section333}.
The real Ginzburg--Landau equation also has steady sech solutions, so
we can find localized solutions of the FCGL equation~\cref{eq:cprs}
in terms of hyperbolic functions. The sech solution
of~\cref{eq:amplitude_real_GL} is
\begin{equation}
  \label{eq:exact_sol_nonzero}
  P(\tilde{X}) = \sqrt{\frac{2 \Gamma_c^2 \Gamma_2}{h_1}}
  \operatorname{sech} 
  \left(\sqrt{\frac{\Gamma_c^2\Gamma_2}{h_2}} \tilde{X} \right)
  e^{i\phi_1},
\end{equation}
where $\phi_1$ is an arbitrary phase and
\begin{align*}
  h_1 & = 3 \left((\rho + 2 \alpha q_c^2) C_r
        + (\nu + v_g q_c + 2 \beta q_c^2) C_i \right), \\
  h_2 &= -2 \left(\rho \alpha + \nu \beta + \tfrac14 v_g^2 + 3 v_g \beta q_c
        + 6 (\alpha^2 + \beta^2) q_c^2 \right),
\end{align*}
and $\Gamma_2$, $h_1$ and $h_2$ must all have the same sign for the
$\operatorname{sech}$ solution to exist.
From~\cref{eq:relationship2} we have
$\bar{Q}(\tilde{X}) = P(\tilde{X}) e^{i\phi}$.

At leading order,
\[
A(X) = \epsilon_2 P(X) e^{iq_cX} =
\sqrt{\frac{2 \Gamma_c (\Gamma - \Gamma_c)}{h_1}} 
\operatorname{sech} \left(\sqrt{\frac{\Gamma_c (\Gamma - \Gamma_c)}{h_2}} 
  X \right) e^{i(q_c X+\phi_1)}
\]
provided $\Gamma < \Gamma_c$, $h_1 < 0$ and $h_2 < 0$. Furthermore,
\cref{eq:A1_B1} and~\cref{eq:relationship2} imply that $\bar{B}(X) =
A(X) e^{i\phi}$.

Finally, recall that in~\cref{eq:sol1}, we wrote the solution to the
original PDE~\cref{eq:rs2} as
\[
U = \epsilon U_1 
= \epsilon \left( A(X,T) e^{ix} + B(X,T) e^{-ix} \right) e^{it}.
\]
Substituting the above formulas for~$A$ and~$\bar{B}$, we find that
\[
U = 2 \epsilon \sqrt{\frac{2 \Gamma_c (\Gamma - \Gamma_c)}{h_1}}
\operatorname{sech} 
\left(\epsilon \, \sqrt{\frac{\Gamma_c (\Gamma - \Gamma_c)}{h_2}} \, x \right)
\cos \Bigl( (1 + \epsilon q_c) x + \tfrac12 \phi + \phi_1 \Bigr) 
e^{i(t - \frac12 \phi)}.
\]
Using \cref{tb:table1}, we return all parameter values to those
used in~\cref{eq:rs2}. Thus, we conclude that the spatially localized
oscillon is given approximately by
\begin{equation}
 \label{eq:exact_as}
 U_{loc}(x,t) = \sqrt{\frac{F_c (F - F_c)}{2h_1^*}}
 \operatorname{sech} \left(\sqrt{\frac{F_c (F - F_c)}{16 h_2^*}} x \right)
 \cos(k_c x + \tfrac12 \phi +\phi_1) e^{i(t - \frac{\phi}{2})},
\end{equation}
where $k_c = 1 + \epsilon q_c$ and $h_1 ^*$ and $h_2 ^*$ are given by:
\begin{align*}
  h_1^* &= 3 \Big( \mu - \alpha + \gamma + 2 \alpha (k_c-1)^2 \Bigr) C_r \\
        &\qquad\qquad + 3 \Bigl( \omega - \beta + \delta - 1 
          - 2 (\beta - 2 \delta) (k_c - 1)
          + 2 \beta (k_c - 1)^2 \Bigr) C_i , \\
  h_2^* &= -2 \alpha (\mu - \alpha + \gamma) 
          -2 \beta( \omega - \beta + \delta - 1) \\
        &\qquad\qquad - 2 (\beta - 2 \delta)^2 
          + 12 \beta (\beta - 2 \delta) (k_c - 1) 
          - 12 (\alpha^2 + \beta^2) (k_c - 1)^2.
\end{align*}
This solution $U_{loc}$ gives an approximate oscillon solution of the
model PDE \cref{eq:rs2} valid in the limit of weak dissipation, weak
detuning, weak forcing, small group velocity, and small amplitude.

\begin{figure}
  \centering
  \includegraphics[width=\linewidth]{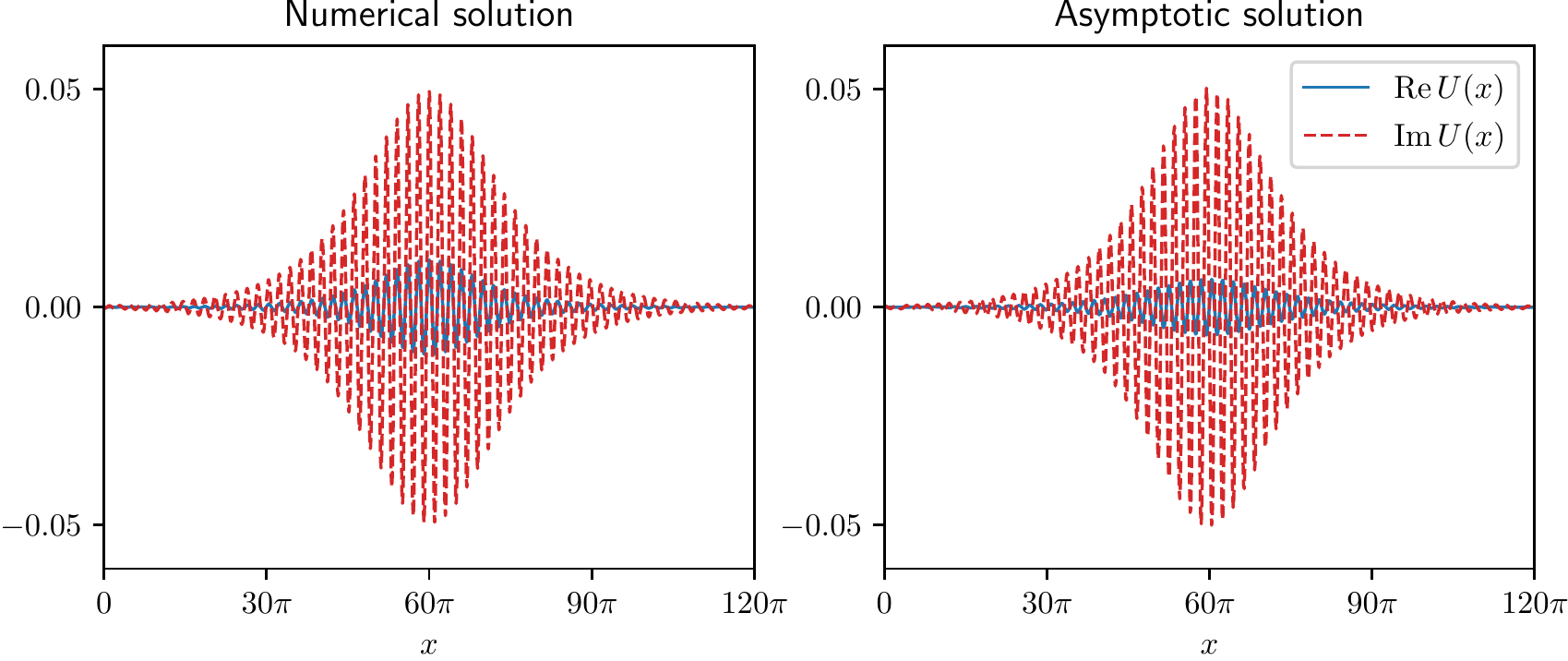}
  \caption{The left panel is the numerical solution of~\cref{eq:rs2},
    reproduced from \cref{fig:five_oscillons}(c). The right
    panel shows the asymptotic solution, given
    in~\cref{eq:exact_as}. These solutions are at $F = 0.073$.}
  \label{fig:compare_with_as}
\end{figure}

In \cref{fig:compare_with_as}, we compare the asymptotic solution
\cref{eq:exact_as} with the localized solution from~\cref{eq:rs2}
which we found numerically in \cref{sec:numerics}.  The similarity
betwen the two is quite striking; the main difference is that the real
part of the asymptotic solution is somewhat smaller than that of the
numerically computed solution, indicating a small error in the
phase~$\phi$.

At this order, we do not find a connection between the position of the
sech envelope and that of the underlying $\cos(k_cx)$ pattern. The
relative position should not be arbitrary, and could presumably be
determined using an asymptotic beyond-all-orders theory~\cite{CK}.

\section{Discussion}
\label{sec:discussion}

In this article, we have shown the existence of oscillons in the
PDE~\cref{eq:rs2}, which was proposed as a pheonomenological model
for the Faraday wave experiments in~\cite{RS}. We first used numerical
simulation, and found that straightforward time-stepping with
carefully chosen parameter values and initial conditions leads to a
stable oscillon solution, as shown in
\cref{fig:ex_pde_chap3_timestepping}. We then turned to
analysis.  Assuming that the damping, detuning and forcing are weak
and that the group velocity and amplitude are small, we reduced the
PDE~\cref{eq:rs2} to the coupled forced complex Ginzburg--Landau
(FCGL) equations~\cref{eq:cprs}. We stress that we do not get a
single FCGL equation with an $\bar{A}$~term, cf.~\cref{eq:fcgl},
which is commonly used as a starting point in discussions of oscillons
in parametrically forced systems~\cite{AT, DL, MS}.  The single FCGL
equation is appropriate when there is a zero-wavenumber
bifurcation~\cite{ANRW} or if the group velocity is zero. However, if
the wavenumber is nonzero (as in Faraday waves) and the group velocity
is nonzero but small, the coupled FCGL equations should be used. The
coupled FCGL equations, and the model PDE, both exhibit snaking
behaviour though the snaking region is very narrow.

Under the further assumption that the strength of the forcing is close
to the onset of instability, we then reduced the coupled FCGL
equations to the subcritical real Ginzburg--Landau
equation~\cref{eq:amplitude_real_GL}. This equation has a sech
solution, which, after undoing the reductions, yields an approximate
expression for the oscillon, cf.~\cref{eq:exact_as}. This expression
agrees well with the oscillon found numerically (see
\cref{fig:compare_with_as}), just as was found in~\cite{ANRW},
where we studied a zero-wavenumber version of this problem.

One special feature of our model PDE~\cref{eq:rs2} is that the linear
terms lead only to positive-frequency oscillations: $U \sim e^{it}$.
With spatial dependence, we have left-travelling and right-travelling
waves, see~\cref{eq:sol1}. In the Faraday wave experiment, as
described by the Zhang--Vi\~{n}als equations~\cite{ZV} or the
Navier--Stokes equations~\cite{SG}, the PDEs are real and so both
positive and negative frequency travelling waves can be found. Topaz
and Silber~\cite{TS} wrote down amplitude equations for these
travelling waves in the context of two-frequency forcing, without long
length scale modulation. In spite of having only positive frequency,
our coupled FCGL equations (with spatial modulation removed) have the
same form as the travelling wave amplitude equations in~\cite{TS}
(after truncation to cubic order). These travelling wave equations
(without modulation terms) can similarly be reduced to standing wave
equations~\cite{PS, TS} with a phase relationship
like~\cref{eq:relationship2} between the complex amplitudes of the
travelling wave components. Therefore, we expect that the fact that
the model PDE~\cref{eq:rs2} has $e^{it}$ dominant should not prevent
oscillons being found by the same mechanism in PDEs that are closer to
the fluid dynamics, because our model PDEs and PDEs for fluid
mechanics lead to the same amplitude equation in the absence of
spatial modulation.

Since \cite{PS, TS} did not include spatial modulations, they did not
have to consider the group velocity. In the present study, we assumed
that the group velocity is small, of the same order as the amplitude
of the solution, in order to make progress. This assumption is
questionable in the context of fluid mechanics. It would be better to
assume that the group velocity is order one, as in~\cite{MKV}. In that
case, the left-travelling wave sees only the average of the
right-travelling wave and vice versa, leading to (nonlocal) averaged
equations. The authors of~\cite{MKV} found spatially uniform and
non-uniform solutions with both simple and complex time dependence,
but did not study spatially localized solutions. Bringing in spatially
localized solutions will be the subject of future work. It is possible
to go directly from the PDE~\cref{eq:rs2} to the real
Ginzburg--Landau equation~\cite{A, RS}, and we expect to be able to do
a simular reduction for the Zhang--Vi\~nals or the Navier--Stokes
equation for fluid mechanics; cf.~\cite{SG, ZV}.

\begin{figure}
  \centering
  \includegraphics[width=\textwidth]{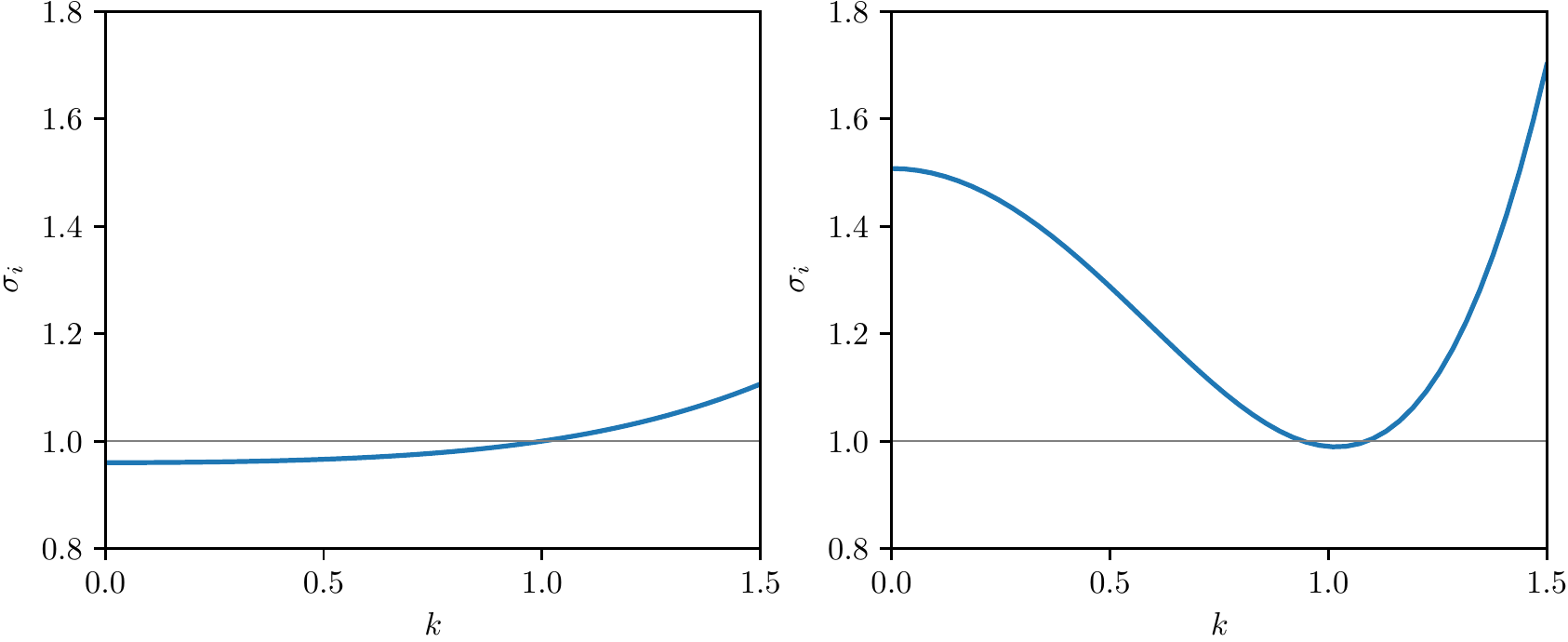}
  \caption{Left panel: The dispersion relation for the PDE model
    \cref{eq:rs2} with $\omega = 0.96$, $\beta = -0.02$ and $\delta =
    0.02$. The frequency $\sigma_i(k)$ is close to~$1$ over a wide
    range of~$k$.  Right panel: Dispersion relation for $\omega =
    1.5075$, $\beta = 1$ and $\delta = 0.4825$.  Now, $\sigma_i(k)$~is
    equal to~$1$ at two distinct wavenumbers.}
  \label{fig:two_wavenumbers_pde}
\end{figure}

\begin{figure}
  \centering
  \includegraphics[width=\linewidth]{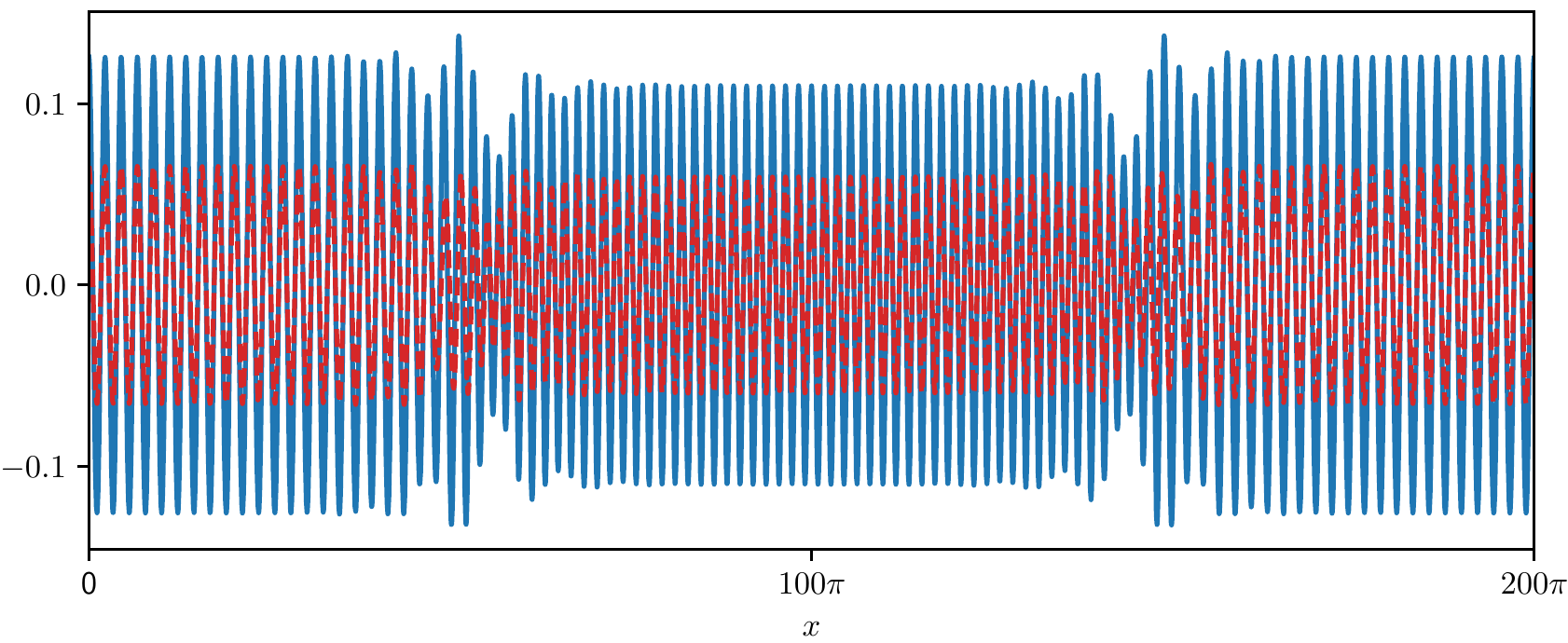}
  \caption{Solution of the PDE model \cref{eq:rs2} with two
    wavenumbers. The parameter values are $\mu = -0.255$,
    $\omega = 1.5075$, $\alpha = -0.5$, $\beta = 1$, $\gamma = -0.25$,
    $\delta = 0.4825$, $C = -1-2.5i$ and $F = 0.15$.}
  \label{fig:two_waves}
\end{figure}

In the model PDE~\cref{eq:rs2}, when the group velocity is small,
waves with a wide range of wavenumbers may be excited.
\Cref{fig:two_wavenumbers_pde} shows two ways in which we can get a
fairly small group velocity. The dispersion curve in the left panel is
shallow; in this case many wavenumbers are close to resonant
($\sigma_i$ is close to~$1$). Another possibility is to have two
resonant wavenumbers around $k=1$, so that $\sigma_i$ is close to a
minimum (where $v_g = 0)$ at $k=1$; see the right panel for an
example. In the latter case, solutions with two nearby wavelengths can
be expected. Indeed, we did observe such solutions in the PDE
model~\cref{eq:rs2}; an example is given in \cref{fig:two_waves}.
These states resemble those found by Bentley~\cite{BA} in an extended
Swift--Hohenberg model, and by Riecke~\cite{RI} in the coupled FCGL
equations with small group velocity in the supercritical case.

\begin{figure}
  \centering
  \includegraphics[width=\textwidth]{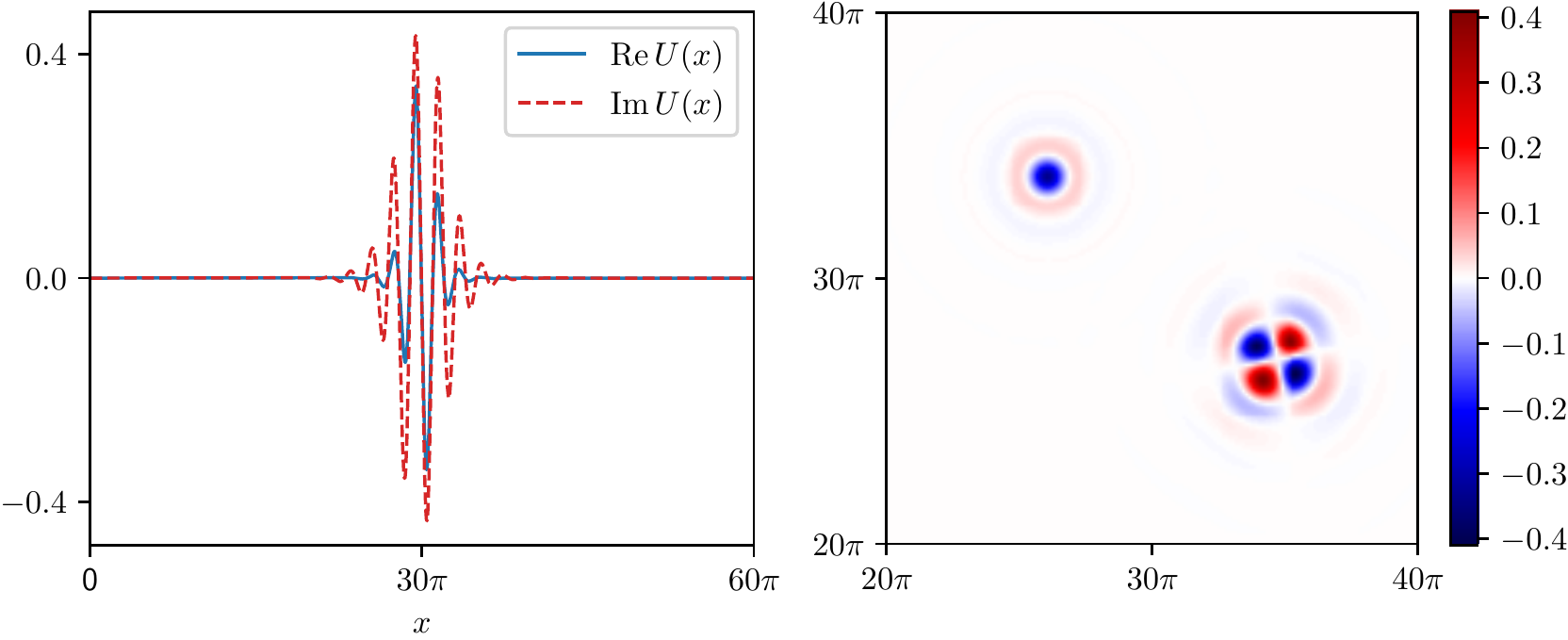} \\
  \caption{Strongly localized oscillons in the PDE
    model~\cref{eq:rs2} in one (left) and two (right) dimensions with
    $\mu = -0.375$, $\omega = 1.99$, $\alpha = -0.5$, $\beta = 1$,
    $\gamma = -0.25$, $\delta = 0.4975$, $C = -1-2.5i$ and $F =
    1.5$. The right panel shows the real part of~$U$ on only part of
    the domain $[0,60\pi] \times [0,60\pi]$.}
  \label{fig:strongly_loczed}
\end{figure}

Finally, we have throughout kept our paramter~$\epsilon$ small
($\epsilon = 0.1$), which is why the oscillons in e.g.\
\cref{fig:five_oscillons} are so broad, in contrast to the oscillons
seen in experiments (see \cref{fig:experiment}). As a preliminary
exploration of increasing~$\epsilon$, we set $\epsilon = 0.5$, and,
after some minor changes to the parameters, we found strongly
localized oscillons in one and two dimensions (see
\cref{fig:strongly_loczed}). As the picture in two dimensions shows,
it is possible for a solution to contain multiple oscillons, which may
or may not be axisymmetric.  Reference~\cite{A} investigates a
related PDE: \cref{eq:rs2}~but with strong damping and with
cubic--quintic (rather than simply cubic) nonlinearity, where the
coefficient of the cubic term has positive real part in order to make
the oscillons more nonlinear. In this case, snaking was found in both
one and two dimensions.

\section*{Acknowledgments} 

We are grateful for interesting discussions with E. Knob\-loch and
K. McQuighan. We also acknowledge financial support from Al Imam
Mohammad Ibn Saud Islamic University.

\end{document}